%% file: EM_probes2.tex
\documentclass[graybox]{svmult}

\usepackage{type1cm}        
\usepackage{makeidx}         
\usepackage{graphicx}        
\usepackage{multicol}        
\usepackage[bottom]{footmisc}
\usepackage{cite}
\usepackage{quoting}

\usepackage{newtxtext}       %
\usepackage[varvw]{newtxmath}       

\newcommand{\kompost}{K{\o}MP{\o}ST\ }

\makeindex             

\begin{document}

\title*{Electromagnetic  Radiation from High-Energy Nuclear Collisions}

\author{Charles Gale}

\institute{Charles Gale  \at    Department of Physics, McGill University, 3600 University street, Montreal QC, Canada H3A 2T8  \email{charles.gale@mcgill.ca}}

%
%
\maketitle

\abstract{
We highlight some of the developments   in the theory and the observation of the electromagnetic radiation, thermal and otherwise,  emitted in relativistic heavy-ion collisions. }

\section{Introduction}
\label{sec:1}
\begin{quotation}
	\begin{flushright}
		\begin{tabular}{l}
			{\em Mama always told me not to look into the eyes of the sun} \\
			{\em But mama, that's where the fun is}\\
		\end{tabular}
	\end{flushright}
	\hspace*{\fill}	-Bruce Springsteen, {\em Blinded By The Light}
\end{quotation}

This article is meant, in the spirit of this volume, as a chronicle of the remarkable progress accomplished  in the theory and in the measurement of electromagnetic radiation emitted from heavy-ion collisions. 
 
Measurements of the electromagnetic radiation emanating from the collision of strongly interacting particles date back to the very early scattering experiments which predate the establishment  of the theory of the nuclear strong interaction, QCD. Owing to the relative weakness of the electromagnetic coupling, $\alpha_{\rm E M}/\alpha_{\rm s} \ll  1$, photons (real and virtual) offered the promise of a clean probe of the reaction dynamics as the mean free paths associated with the electromagnetic interaction exceed typical hadronic length scales. In the modern era ushered  in by the availability of high energy hadronic beams, the immense potential of electromagnetic observables became clear, and their status as natural complements to measurements of hadrons was confidently established. To make the discussion specific, the photon production cross section in proton-proton collisions, for example, is calculable in perturbative QCD (pQCD), with the help of factorization  \cite{Field:1989uq}. Consequently, early measurements of electromagnetic radiation at CERN (ISR) and at Fermilab where crucial in the confirmation  of the parton model, of QCD itself, and of its perturbative treatment \cite{Drell:1970wh,Owens:1986mp}. 

To repeat, the basic feature which makes real and virtual photons so appealing in  studies of QCD is that  they suffer little final state interactions and therefore decouple from the strongly interacting system. They are therefore good probes of the  conditions local to their production site, assuming their formation mechanism  is under theoretical control. This is also the reason why electromagnetic measurements have accompanied the development of the field now known as relativistic heavy-ion collisions which concerns itself with the formation, study, and characterization of the quark-gluon plasma, an exotic state of the fundamental constituents of QCD and which is also the central theme of the present volume. 

\begin{figure}[htbp]
	\begin{center}
		\includegraphics[scale=0.3]{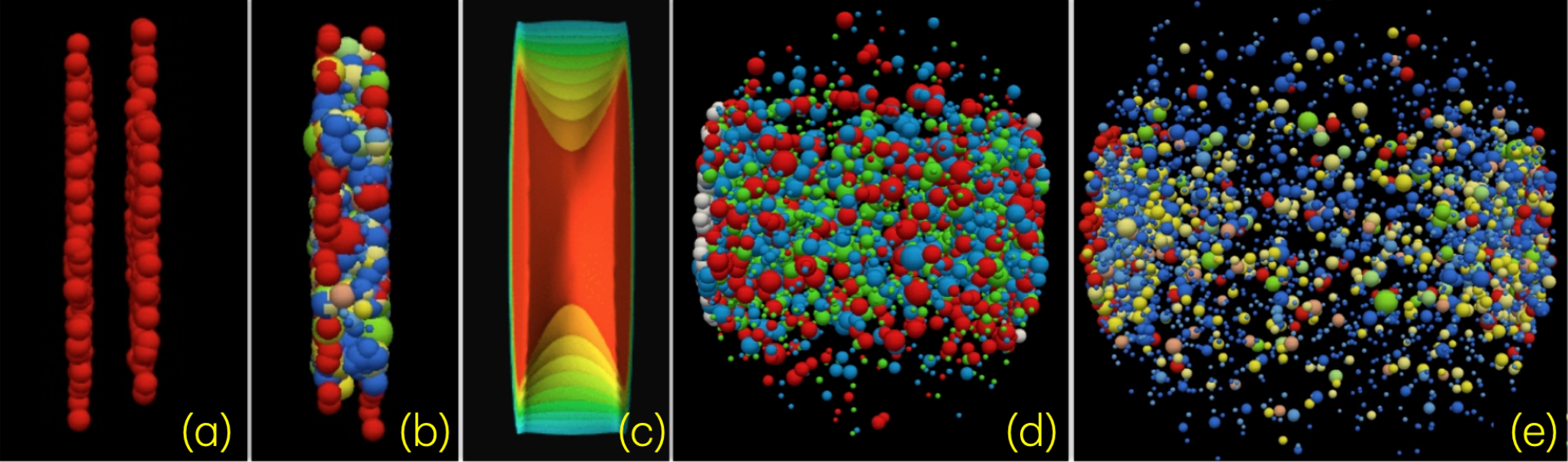}
	\end{center}
	\caption{The many stages of a heavy-ion collision: (a) Initial state; (b) pre-hydro phase; (c) fluid-dynamical evolution; (d) Hadronization; (e) Final-state hadrons eventually free-streaming to the detectors.   These images are not from a single end-to-end  calculation, but are simply meant to illustrate the different stages. The pictures have been assembled and collected from different sources which include hydrodynamical simulations \cite{Schenke:2010nt} and UrQMD calculations \cite{Bleicher:1999xi}. }
	\label{fig-stages}       
\end{figure}

Our current understanding of global  heavy-ion reaction dynamics is illustrated in Fig.~\ref{fig-stages} which shows the different eras which exist in those high-energy scattering events. The need to theoretically address strongly interacting matter over large scales of length and time has driven the development of the multistage approaches currently used to  model the evolving system. In heavy-ion collisions, real and  virtual photons -- the latter will decay into lepton pairs -- will be produced at all the different stages of the collision process: this signal can therefore serve to validate multistage models that have been tuned to a variety of hadronic observables as well as act as a tomographic probe of the system.  Relativistic heavy-ion physics can rightly claim to be a strong proponent  of multimessenger  science.

 In the rest of this paper, we discuss the techniques needed and used to calculate the generation of electromagnetic radiation from hadronic collisions, concentrating on signals that are thermal and thermal-like. 

\section{Electromagnetic radiation from systems in equilibrium}
\label{sec:2}
As photons leave the reaction site unscathed, they potentially carry pristine information about the local conditions at their creation site. For example, high energy photons can report on 
the early instants of the collision, an era hardly accessible to hadronic signals, with the possible exception of QCD jets and of heavy quarks. Both of those however undergo in-medium final state interactions. 

Some of the early exploratory  calculations devoted to estimates of the temperature of the quark-gluon plasma have proposed measuring the spectrum of lepton pairs of invariant mass $M$ in the region  $m_\phi < M \lesssim m_{J/\psi}$ and extracting an effective temperature from its slope,  in analogy with black body radiation \cite{Shuryak:1978ij,Kajantie:1981wg,Domokos:1980ba,Kajantie:1986dh}. This basic philosophy -- using electromagnetic radiation to characterize the QGP --  is followed to this day, but with some of the additional sophistication described  in this paper. 

In a strongly interacting medium at temperature $T = 1/\beta$, a starting point to the computation of thermal electromagnetic emission rates is the transition rate between two states, an initial state  $i$ and a final state  $f$,  which can be written as $R_{f i}= \frac{\left| S_{f i}\right|^2}{\tau V}$ with $\tau V$ being the proper four-volume. To leading order in the electromagnetic interaction, the interaction matrix element written in terms of the electromagnetic current operator  $J^\mu$ coupled to the photon field $A^\mu$ is $S_{f i} = \langle f | \int\!\! d^4 x J^\mu (x) A_\mu (x) |  i  \rangle$.  The vector field may then be written as $A^\mu (x) = \epsilon^\mu \left( e^{i k\cdot x} + e^{-i k\cdot x}\right)/\sqrt{2 \omega V}$, where $\epsilon^\mu$ is a polarization vector and $k^\mu = \left( \omega, {\bf k}\right)$.  The photon  rate is
\begin{eqnarray}
R_{f i} = - \frac{g^{\mu \nu}}{2 \omega V} (2 \pi)^4 \left[\delta\left(p_i + k - p_f\right) + \delta \left(p_i - k - p_f\right)\right] \langle f |J_\mu (0) | i \rangle \langle i | J_\nu (0) | f \rangle \nonumber \\
\end{eqnarray}
where one delta function deals with the absorption process, the other with emission.

 Using the equation of motion (or Maxwell's equations) and the Schwinger-Dyson equation   leads to 
the expressions for the differential emission rate $\Gamma $ (electromagnetic quanta per unit four-volume) of photons and dileptons \cite{Feinberg:1976ua,McLerran:1984ay,Weldon:1990iw,Gale:1990pn,Kapusta:2007xjq}: 
\begin{eqnarray}
\omega	\frac{d \Gamma_\gamma}{d^4 k} &=& - \frac{g^{\mu \nu}}{\left(2 \pi\right)^3} {\rm Im} \,\Pi_{\mu \nu} (\omega, {\bf k}) \frac{1}{e^{\beta \omega}-1}\ \ {\rm (photons)}\nonumber \\
E_+ E_- \frac{d\Gamma_{\ell \bar \ell}}{d^3 p_+ d^3 p_-} &=& \frac{2 e^2}{\left( 2 \pi \right)^6} \frac{1}{k^4}  L^{\mu \nu} \, {\rm Im} \Pi_{\mu \nu} (\omega, {\bf k}) \frac{1}{e^{\beta \omega} - 1}\ \ {\rm (dileptons)}
\label{rate_eqs}
\end{eqnarray}
The lepton tensor is (for leptons with rest mass $m_\ell$) $L^{\mu \nu} = p_+^\mu p_-^\nu + p_+^\nu p_-^\mu - g^{\mu \nu}\left( p_+ \cdot p_- + m_\ell^2\right)$. The photon (dilepton) four-momentum  $k^\mu $ is  light-like for real photons and timelike for dileptons. In these equations, $\Pi^{\mu \nu}$ is the finite-temperature retarded in-medium photon self-energy; it contains the information on how the medium affects the particles eventually assembling into real or virtual photons. Importantly, these equations are valid up to first order in the electromagnetic coupling, but are {\em exact} in the strong coupling \cite{Gale:1990pn}. This formalism is therefore directly amenable to non-perturbative  field theoretical calculations, where the photon self-energy can be written as a current-current correlator  (the electromagnetic current)  \cite{Gale:1990pn} obtainable by analytically continuing the Euclidean correlator evaluated non-perturbatively on the lattice. For a non-Abelian gauge theory like QCD this is a technically challenging procedure \cite{Ghiglieri:2016tvj}, but  recent progress has been reported  \cite{Ce:2022fot,Ali:2024xae}.

An alternative approach to the evaluation of electromagnetic radiation production rates is provided by the fact that taking the imaginary part of the photon self-energy is tantamount to putting internal virtual particles on-shell, enabling in principle a formalism based on kinetic theory \cite{Weldon:1983jn,McLerran:1984ay,Majumder:2001iy}. This correspondence is most readily seen in the case of dilepton production where the ``simplest'' (in terms of loop topology) diagram of the (virtual) photon self-energy is a single fermion loop:
 \begin{figure}[h!]
 	\begin{center}
 		 	\vspace*{-0.5cm}
 		\includegraphics[scale=0.15]{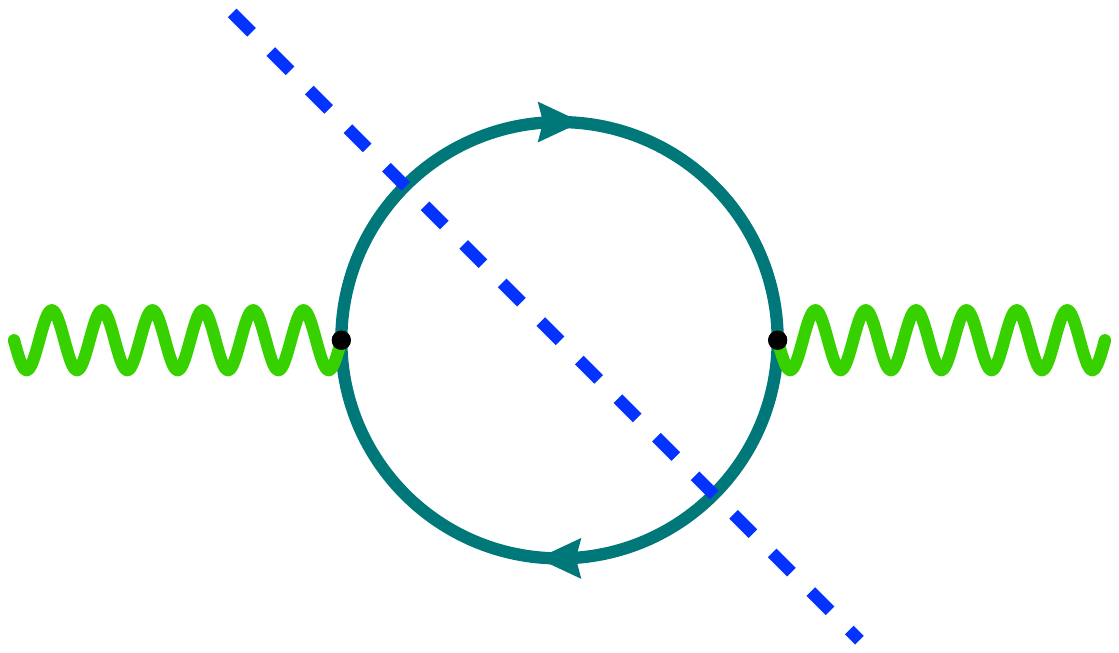}
 	\end{center}
 	%
 	\label{Loop}       
 \end{figure}
\vspace*{-2.9cm}
where the dashed line indicates the taking of the imaginary part or colloquially ``cuttting the diagram''.  The cut diagram then reveals the $q \bar q \to \gamma^*$ process \cite{Gale:1990pn}, often calculated using kinetic theory. 

In this language, the photon production rate from the reaction $1 + 2 \to 3 + \gamma$ is
\begin{eqnarray}
	\Gamma_\gamma &=& {\cal N} \int \frac{d^3 p_1}{2 E_1 \left( 2 \pi\right)^3} \frac{d^3 p_2}{2 E_2 \left( 2 \pi\right)^3} f_1 (E_1) f_2 (E_2) \left( 2 \pi\right)^4 \delta\left( p_1^\mu + p_2^\mu - p_3^\mu - k^\mu\right) \nonumber \\
&&	\times \left|{\cal M}\right|^2 \frac{d^3 p_3}{2 E_3 \left( 2 \pi\right)^3}\frac{d^3 k}{2 \omega \left( 2 \pi\right)^3}\left[ 1 \pm f_3 (E_3)\right]
\end{eqnarray}
where ${\cal N}$ is an overall degeneracy factor, the $f$'s are Fermi-Dirac or Bose-Einstein distribution functions as appropriate, and the matrix element for this interaction is ${\cal M}$.  Those two  approaches have their own advantages and disadvantages and are often used as being  complementary, even if going from one formalism to the other can hide some subtleties and care is needed \cite{Wong:2000hq}.

The historical development of thermal emission rates from strongly interacting matter as that created in heavy-ion collisions is an interesting story in itself.  Calculations of photons and dileptons from the hot QGP involve the interactions between gluons and quarks in a finite-temperature medium. Early kinetic theory calculations provided estimates of photon and dilepton emission from interacting partons \cite{Shuryak:1978ij,Kajantie:1981wg,Sinha:1983jm,Hwa:1985xg,Staadt:1985uc}. Writing the production rate of real photons  in terms of the photon self-energy with Eq. (\ref{rate_eqs}) in mind, the processes at leading order in $\alpha_s$:  $q g \to q \gamma$ and $q \bar{q} \to g \gamma$ appear at two loops in the loop expansion of the photon self-energy. If the virtual quarks exchanged in those reactions are massless, the emission rate suffers from a logarithmic collinear singularity, cured however by the generation of an in-medium effective fermion mass calculated using  the Hard Thermal Loops (HTL) resummation \cite{Kapusta:1991qp,Baier:1991em}. The formalism of HTL was  established earlier and used to famously address the manifest and vexing gauge dependence that plagued calculations of the gluon damping rate in finite temperature QCD\footnote{See Ref. [3] in Ref. \cite{Braaten:1989mz}.}. 
\begin{figure}[htbp]
	\begin{center}
		\vspace*{-0.5cm}
		\includegraphics[scale=0.35]{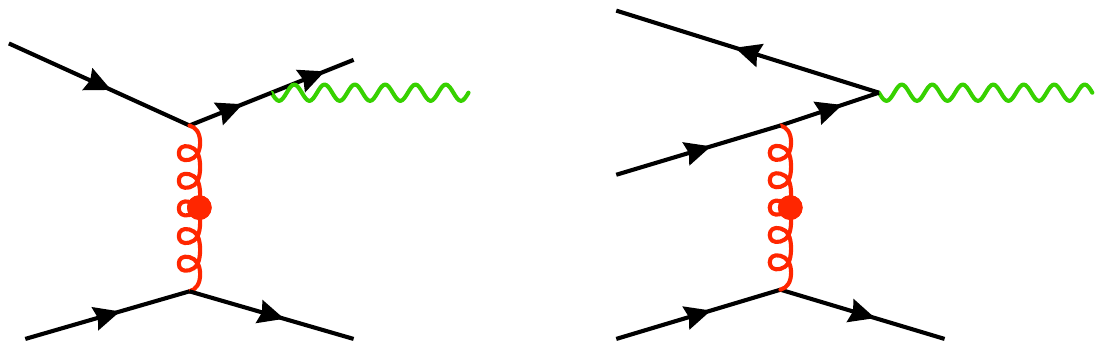}
		\vspace*{-7cm}
	\end{center}
	\caption{Two photon-producing processes that come from taking the imaginary part of a two-loop photon self-energy. They are (left) bremsstrahlung from a quark (or anti-quark), and (right)  $q {\bar q}$ annihilation with scattering. The gluon propagator is HTL-corrected (see main text).}
	\label{fig-brem}       
\end{figure}

Further explorations in the topology of the photon self-energy lead to evaluations of the parton bremsstrahlung contribution to the production of real photons. Using an effective theory based on the resummation of HTLs, this channel enters at the two-loop level in that effective theory and taking the imaginary parts produces the processes shown in Fig. \ref{fig-brem}. Even though the naive parametric power counting of the  photon self-energy would have those contributions $\sim g^4$, where $g$ is the strong coupling constant, it was found that collinear divergences formally promote the photon rate to $\sim g^2$ \cite{Aurenche:1998nw}. The solution to the puzzling question  of how to uncover and consistently sum up all the contributions at a given order of the perturbation expansion  was provided with the inclusion of photons generated by the Landau-Pomeranchuk-Migdal effect (LPM), which limits the coherence length of the emitted radiation. The first photon production rates complete at leading order (LO) in the strong coupling were presented in Ref. \cite{Arnold:2001ms}, and results at NLO were reported later in Ref. \cite{Ghiglieri:2013gia}\footnote{For values of $\alpha_s$ relevant for the phenomenology discussed here, NLO + LO results show an enhancement of $\sim 20 \%$ over LO, owing to some partial cancellations in the NLO contributions.}. 

At lower temperatures, the photon rates calculated with the hadronic degrees of freedom relevant to stages (d) and (e) of Fig. \ref{fig-stages}, have been calculated using interactions modeled with effective chiral hadronic Lagrangians  in Refs. \cite{Meissner:1987ge,Kapusta:1991qp,Song:1993ae,Turbide:2003si,Heffernan:2014mla,Holt:2015cda}. 
Calculations of dilepton production involving composite hadrons also rely on Eq. (\ref{rate_eqs}), but the virtual nature of the photon field reveals the vector meson spectral distribution, owing to the coupling of the photon field with vector meson fields: Vector Meson Dominance \cite{Gale:1990pn,Rapp:1999ej}.

\subsection{Electromagnetic emission out of equilibrium}

Up to this point in our discussion, the medium emitting photon and dilepton  radiation was assumed  in equilibrium; thermal and chemical. Those assumptions were also made  in most of the early estimates of the electromagnetic emissivity of strongly interacting media. However, the modern understanding of the evolution of relativistic heavy-ion reactions is that the strongly interacting medium is never in total equilibrium in any formal sense.  Referring back to Fig. \ref{fig-stages}, some electromagnetic signal will also be emitted in the stage corresponding to that depicted in (b), and the distinction from a signal coming from a source closer to equilibrium  is not possible. Those sources have to be addressed consistently, ideally within a single approach. The theory of radiation emitted out of equilibrium already has a rich history \cite{Michler:2012mg,Schenke:2005ry}, but it is also fair to write that a systematic treatment of the pre-hydrodynamics era in relativistic heavy-ion collisions is still being developed. The following is but a summary of some of the recent efforts. 

To fix ideas, we first consider the emission of real photons; of course, virtual photons are also emitted out of equilibrium. Before discussing the different models addressing photon emission out of equilibrium and just prior to the hydrodynamics phase, it is instructive to discuss the most extreme case: electromagnetic radiation from the very first nucleon-nucleon collisions. Those photons will appear as an irreducible background to any of the other photon sources under consideration, and will typically occupy the highest $p_T$ part of the spectrum.  As alluded to previously, those are calculable with pQCD, using a procedure that is however not free from uncertainties which will grow as $p_T$ decreases: they are associated with scale uncertainties in the nuclear parton density functions (nPDF) \cite{Aurenche:2006vj,Arleo:2011gc,Klasen:2013mga,Paakkinen:2017jpo} and with the relatively poorly known photon fragmentation function (especially from gluons) \cite{Klasen:2014xfa}. This is why the pQCD contributions in the low momentum part $p_T~\approx~1$ GeV/c of the thermal photon window are mostly extrapolations \cite{Paquet:2015lta} which emphasize the importance of direct measurements of low transverse momentum photons in pp and pA collisions.

Just prior to, but also in  the fluid phase, non-equilibrium aspects will manifest themselves in different ways. In the recent couple of decades, the relativistic heavy-ion program pursuing the characterization of the QGP has revealed the remarkable success of viscous fluid dynamical modeling \cite{Gale:2013da}. By comparing with hadronic data, it  became apparent that the results of the early calculations based on inviscid hydrodynamics could be brought closer to measurements with the inclusion of transport coefficients such as shear and bulk viscosity  \cite{Gale:2013da};  the electromagnetic emissivity evaluations also needed to be adapted consequently.  

Deviations from kinetic equilibrium can be obtained by considering the linearized Boltzmann equation and evaluating the deviations ($\delta f$) of the distribution functions ($f = f_0 + \delta f$) from their equilibrium form ($f_0$), using techniques \cite{Denicol} pioneered by Chapman and Enskog \cite{Chapman}, and by Grad \cite{Grad}. Using that approach, the photoproduction channels in the QGP and in the confined hadronic sector involving $2 \to 2$ kinematics  can be corrected for the presence of shear and bulk viscosity \cite{Dion:2011pp} demanded for consistency with the fluid dynamical evolution \cite{HeinzSchenke}.  The electromagnetic emission rates derived thusly can then be implemented in parallel with the fluid dynamical evolution \cite{Paquet:2015lta}. Part of this general framework for correcting the photon emission rates is discussed in Ref. \cite{Dion:2011pp}, and a discussion of the $2 \to 2$ photon-producing contributions in the QGP (Compton and quark-anti-quark annihilation) appears in Ref. \cite{Shen:2014nfa}. 
However, note that not all the photoproduction channels currently used in phenomenology include such out-of-equilibrium corrections: an example is photon production related to the LPM effect. 

It is known that formally, a field-theoretic formulation appropriate for non-equilibrium contributions is based on the real-time formalism as expressed in the ``1/2'' \cite{Schwinger1961,Keldysh:1964ud} and ``r/a'' bases\footnote{The numbers (1/2) label two branches in the complex time plane and (r/a) stands for ``retarded'' and ``advanced''.} \cite{Keldysh:1964ud}. In that language, the thermal photon spectrum is \cite{Serreau:2003wr}
\begin{eqnarray}
	\omega \frac{d R}{d^3 k} = \frac{i}{2 \left( 2 \pi\right)^3} \Pi_{1 2 \mu}^{\mu}\ , 
\end{eqnarray}
where the photon polarization tensor is no longer the retarded one, but contains the  ``1'' and ``2'' vertices \cite{Keldysh:1964ud}. 
Building on this, the effect of the anisotropy in momentum space generated for instance  by the presence of a shear pressure tensor $\pi^{\alpha \beta}$ in photon production in general -- including LPM -- has been elaborated in Ref. \cite{Hauksson:2017udm}, paving the way for an eventual holistic treatment of out-of-equilibrium effects, even if the presence of early medium instabilities requires some care \cite{Hauksson:2020wsm}.

Studies of photon production obtained by solving  some form of kinetic transport theory in the partonic phase prior to ``hydrodynamization'' have also been performed \cite{Chiu:2012ij,Linnyk:2015tha,Greif:2016jeb,Vovchenko:2016mtf,Berges:2017fsa,Oliva:2017pri,Garcia-Montero:2023lrd}. 
The kinetic theory approaches aim to solve the Boltzmann equation and therefore to also capture the non-equilibrium aspects of the space-time evolution. However, in such simulations the inclusion of the LPM effect -- which is due to quantum interference between multiple scatterings -- has been more challenging, and has driven the need for different approximations \cite{Cleymans:1992kb,Greif:2016jeb,Oliva:2017pri}.

 At least in principle, the non-equilibrium behavior of a collection of partons can be assessed by solving a set of coupled Boltzmann equations which contain all of the relevant scattering processes. In \kompost\kern-1ex, as introduced and defined in Refs.  \cite{Kurkela:2018vqr,Kurkela:2018wud}, the  kinetic theory of Ref.  \cite{Arnold:2002zm} is used to construct an energy-momentum tensor $T^{\mu \nu}$ which is coarse-grained and acts as an average background over which fluctuations are propagated using linear response theory. Owing to some robust scaling properties of the Green's functions used in the linear response regime \cite{Kurkela:2018vqr}, this alternative approach to the full solutions of the Boltzmann equation offers some practical advantages. With this approach at hand, one can also evaluate the contribution of the non-equilibrium regime to electromagnetic radiation production using a procedure detailed in Ref. \cite{Gale:2021emg}. The energy-momentum tensor generated by \kompost is decomposed as it is in the hydrodynamics phase, and the energy density, flow velocity, and viscosity contributions are identified. Then, a cell-by-cell effective temperature is extracted at each time step of the  dynamical evolution  using the QCD equation of state \cite{Bazavov_2014}. Even though this procedure is approximate, it enables continuity of the electromagnetic emissivity between the fluid dynamical epoch and that preceding it.  In a given practical implementation, the \kompost stage is inserted between an IP-Glasma initial state \cite{Schenke:2012wb} which is itself derived from the Color Glass Condensate picture of high-energy hadronic scattering, and the viscous hydrodynamics evolution \cite{Gale:2013da}. 
 \begin{figure}[htbp]
 		\begin{center}
 			\includegraphics[width=0.7\textwidth]{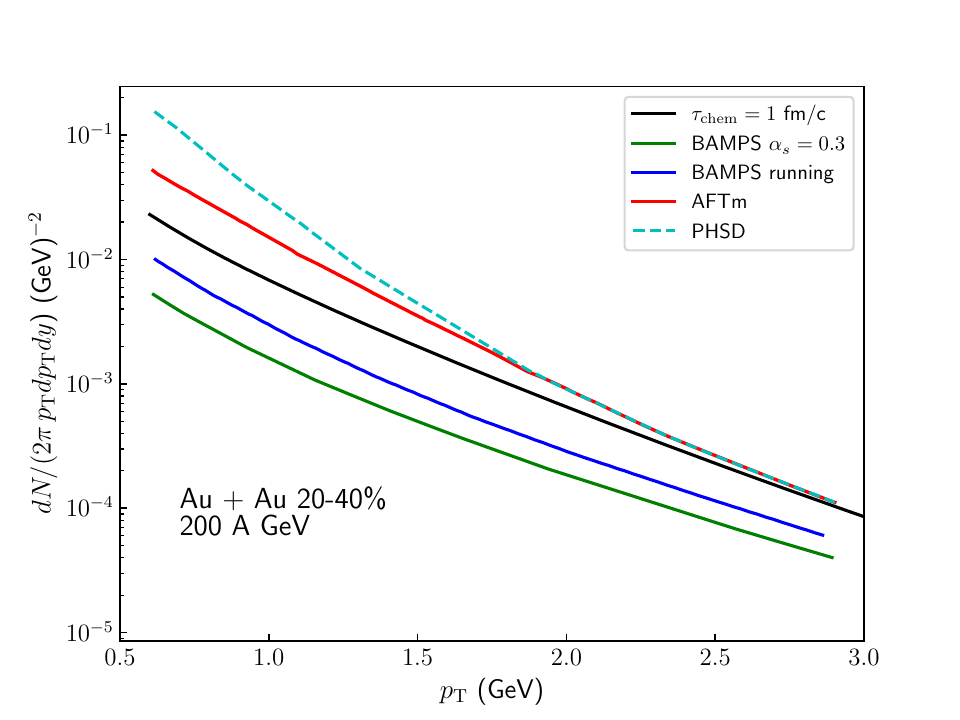}
 			\caption{The invariant photon spectrum plotted as a function of photon transverse momentum at midrapidity, for conditions corresponding to Au + Au collisions at RHIC, in the 20-40\% centrality window. The transport theory results are those obtained with: BAMPS with fixed and running strong interaction coupling; the Abelian Flux Tube model (AFTm); the Parton Hadron String Dynamics (PHSD). The results from those models were adapted from Fig. 8 in \cite{Oliva:2017pri}. The photon results obtained using \kompost\kern-1ex, as in Ref. \cite{Gale:2021emg}, are also shown and labeled $\tau_{\rm chem} = 1$ fm/c (see main text). }
 			 	\label{fig-preq_comp}       
 		\end{center}
 \end{figure}

 Clearly, the exact  chemical composition of the initial state will also influence the radiation  of photons, which require some fermions in the initial state. The fermion content in many approaches is evolved dynamically from an initial  gluon-dominated many-body wave function. This makes the measurement of photons -- both real and virtual -- even more interesting as they can report not only on the effective temperature  of the medium, but also on its early chemical content. 
 
 With all of those aspects  in mind and several of them still in flux,  it is revealing to compare the spectra of real photons obtained with the non-equilibrium approaches detailed in Refs. \cite{Linnyk:2015tha,Greif:2016jeb,Oliva:2017pri,Gale:2021emg} and shown in Fig. \ref{fig-preq_comp}. In each of those calculations, the initial state was treated differently. The AFTm (Abelian Flux Tube model) sets up an initial chromo-electric field which decays and populates the fermion distributions \cite{Oliva:2017pri}. In PHSD (Parton Hadron String Dynamics) \cite{Linnyk:2015tha} it is string breaking which initializes  the partons, which are evolved with the relativistic transport theory. The BAMPS (Boltzmann Approach to Multiparton Scaterrings) simulations \cite{Greif:2016jeb} are based on a gluon-dominated initial state, as does the multistage calculation involving \kompost discussed in \cite{Gale:2021emg}, but the evolution times of the classical Yang-Mills equations are a parameter of the model. In BAMPS, the   gluon Wigner distributions are sampled from the solution of the classical Yang-Mills and then propagated in the partonic cascade.  
 In the multistage model the energy-momentum tensor is passed from IP-Glasma to \kompost and then fed to the hydrodynamic evolution. Clearly all the approaches discussed so far  -- those with results highlighted in Fig. \ref{fig-preq_comp} -- differ in the details of their respective implementation even though some of the basic principles are common.  The size of the non-equilibrium photon production as compared with the more conventional thermal source can be assessed by examining, for instance,  Figure \ref{fig-spectrum}. 
 
 It is still early to come to a definite picture: the  theory of non-equilibrium electromagnetic radiation in heavy-ion collisions has experienced rapid growth in the last few years and features many results in addition to those shown in Fig. \ref{fig-preq_comp}. This topic is now at the point where more detailed comparisons between models would serve the field well, and new data with reduced uncertainties could also make it possible for some of the model parameters to be incorporated in Bayesian studies to follow those of the current generation  \cite{JETSCAPE:2020shq,Nijs:2020roc,Heffernan:2023gye}. In the current state of knowledge and for conditions (system and energy) relevant to RHIC as shown on the figure, the different model assumptions generate non-equilibrium photon spectra which can easily differ by an order of magnitude for low $p_{\rm T} \approx 1$ GeV, and roughly half that at $p_{\rm T} \approx$ 3 GeV.  It is therefore important to compare the pre-equilibrium signal to that generated from other stages of the collision, and with the available data. 
 
 \subsection{Net photon yields}
 
In heavy-ion collisions, direct real photons have been measured by several collaborations;  in the context  of the search for the quark gluon plasma (QGP) earlier measurements were conducted by WA93 \cite{WA93:1998uhf} and WA98 \cite{WA98:1999rbo} at the CERN SPS, followed by STAR \cite{Johnson:2002xj} and PHENIX \cite{Reygers:2002kc} at the Relativistic Heavy-Ion Collider (RHIC), and by ALICE \cite{Wilde:2012wc} at the Large Hadron Collider (LHC). We concentrate on  data from RHIC and the LHC and will show results obtained with  the multistage approach of Ref. \cite{Gale:2021emg}, which  begins with IP-Glasma followed by \kompost and then viscous fluid dynamics. The late hadronic state corresponding to a combination of the phases (d) and (e) of Fig. \ref{fig-stages} of course also emits real and virtual photons ({\it e.g.} $\pi \rho \to \pi \gamma$,  etc) \ \cite{Gotz:2021dco,Gale:2021emg}. These next results are obtained using all the currently available direct photon rates integrated with a relativistic viscous fluid dynamical calculation tuned to reproduce hadronic data. 

\begin{figure}[htbp]
 		\includegraphics[width=5.5cm]{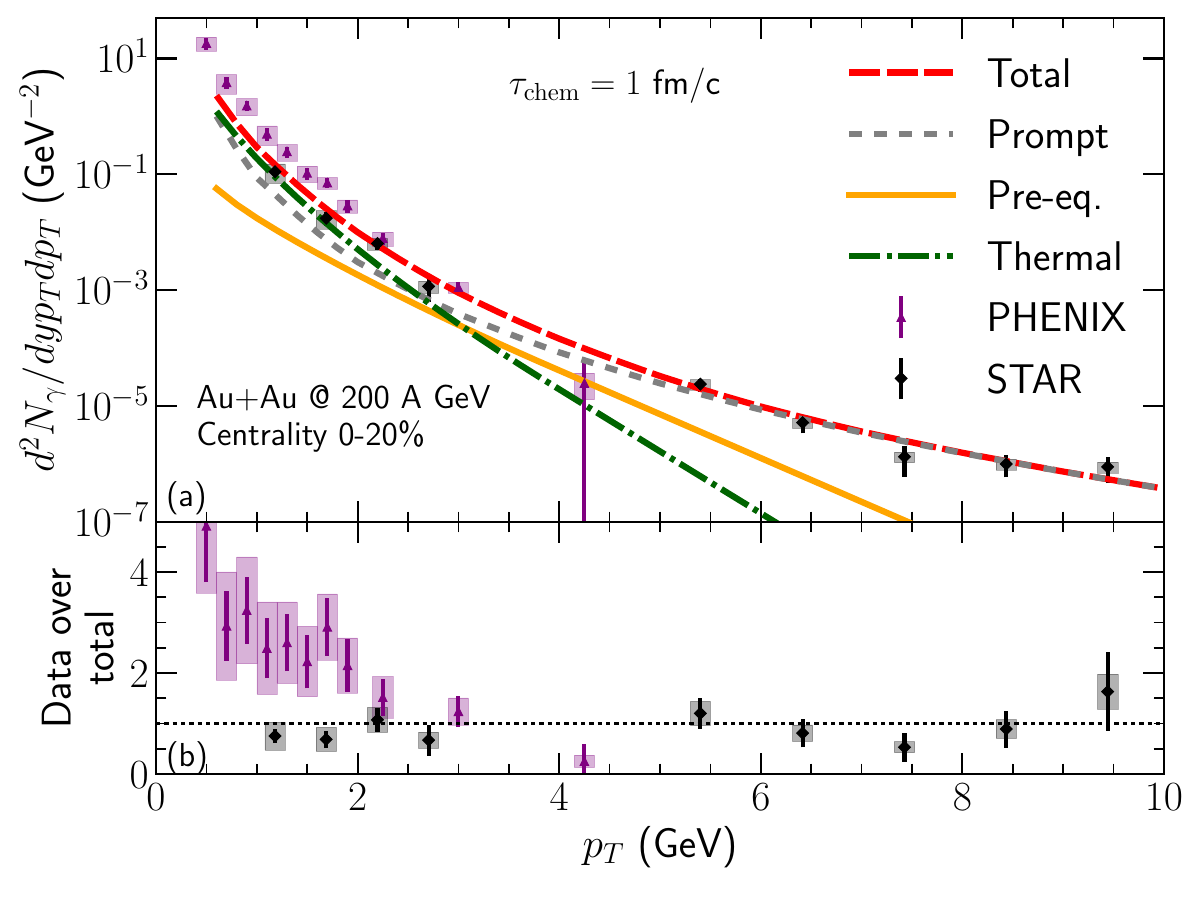}
 		\hspace*{0.2cm}
 		\vspace*{-0.2cm}
 		\includegraphics[width=5.5cm]{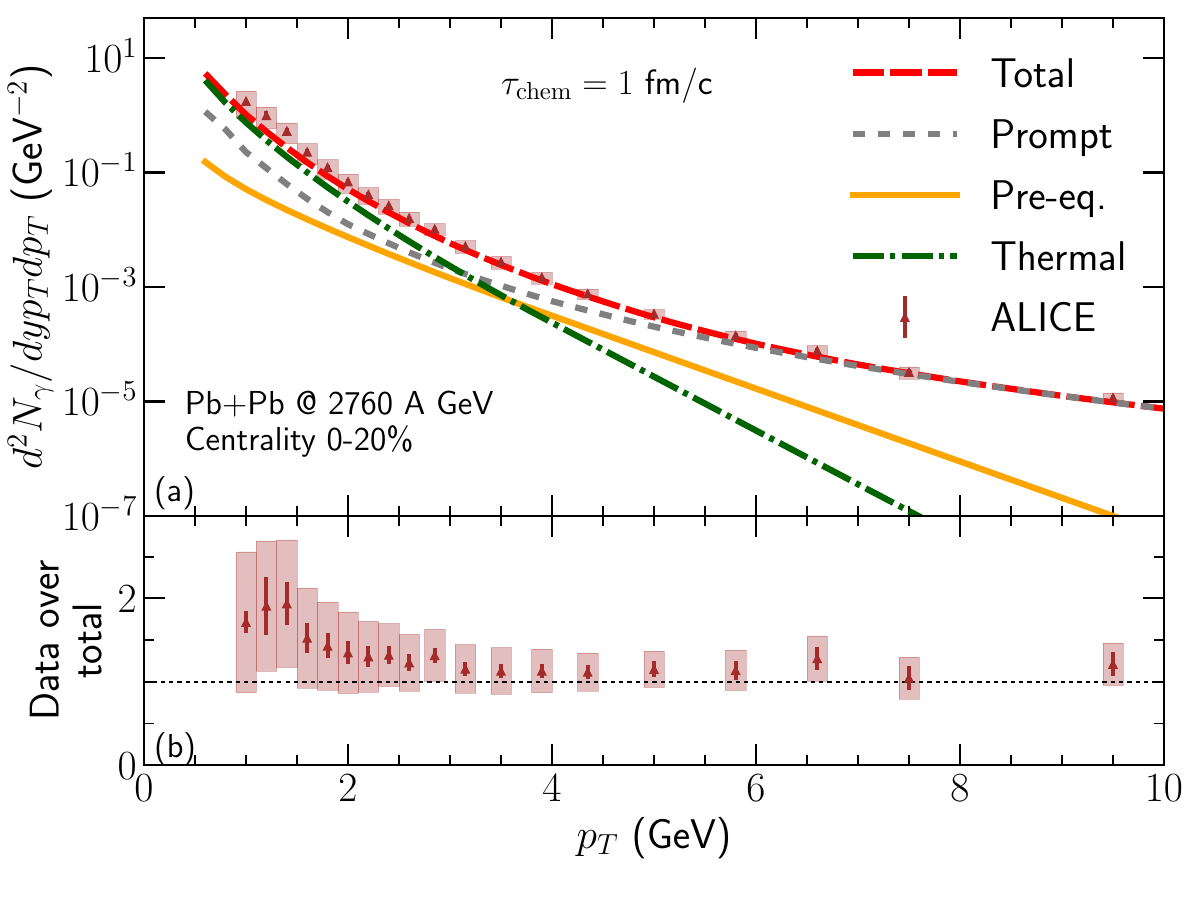}
 	\caption{
 		The yield distribution (top panels) of real photons shown with the appropriate experimental data measured in a $0 - 20\%$ centrality window. The contributions of the different processes discussed in the text are shown separately. (Left panel) Photons from Au + Au collisions at 200 $A$ GeV. The data are from the PHENIX \cite{PHENIX:2014nkk} and STAR \cite{STAR:2016use} Collaborations. (Right panel)   Photons from Pb + Pb collisions at 2760 $A$ GeV. The data are from the ALICE Collaboration \cite{ALICE:2015xmh}. The bottom panels show the ratio of data over the calculated result. Figure adapted from \cite{Gale:2021emg}. }
 	\label{fig-spectrum}       
 \end{figure}

 Figure \ref{fig-spectrum} displays direct photon yields\footnote{The photons from decays of hadrons are measured and subtracted in order to obtain the direct signal. }  and contains  data gathered by the PHENIX and STAR collaborations in collisions of Au + Au at 200 $A$ GeV, at RHIC, and by the ALICE collaboration in collisions of Pb + Pb at 2760 $A$ GeV at the LHC. The parameter $\tau_{\rm chem}$ is a chemical relaxation time, and its significance and value are discussed in Ref. \cite{Gale:2021emg}.  
 The sum of direct real photons (long dashes) is further decomposed into channels which include:  photons from the first-impact collisions calculated with pQCD  (``prompt'', short dashes); pre-hydrodynamics photons (``pre-equilibrium'', solid); thermal photons from the QGP and from the ensemble of composite hadrons (dot-dash). One observes that at RHIC, the pre-equilibrium, thermal, and pQCD photon contributions are comparable at  $p_T \approx 3$ GeV, and at the LHC, for $p_T \approx 3.5$ GeV: the pre-equilibrium contribution represents $\gtrsim 30\%$ of the net photon yield  and thus has a promising future as an observable signal. At the LHC energy shown in Fig. \ref{fig-spectrum}, the model and data agree within statistical uncertainties.  At RHIC, there is unfortunately a long-standing difference between the photon spectra measured by STAR and PHENIX, which is so far not fully understood  \cite{David:2019wpt}.
 \begin{figure}[bhtp]
 	\begin{center}
 		\includegraphics[width=0.63\textwidth]{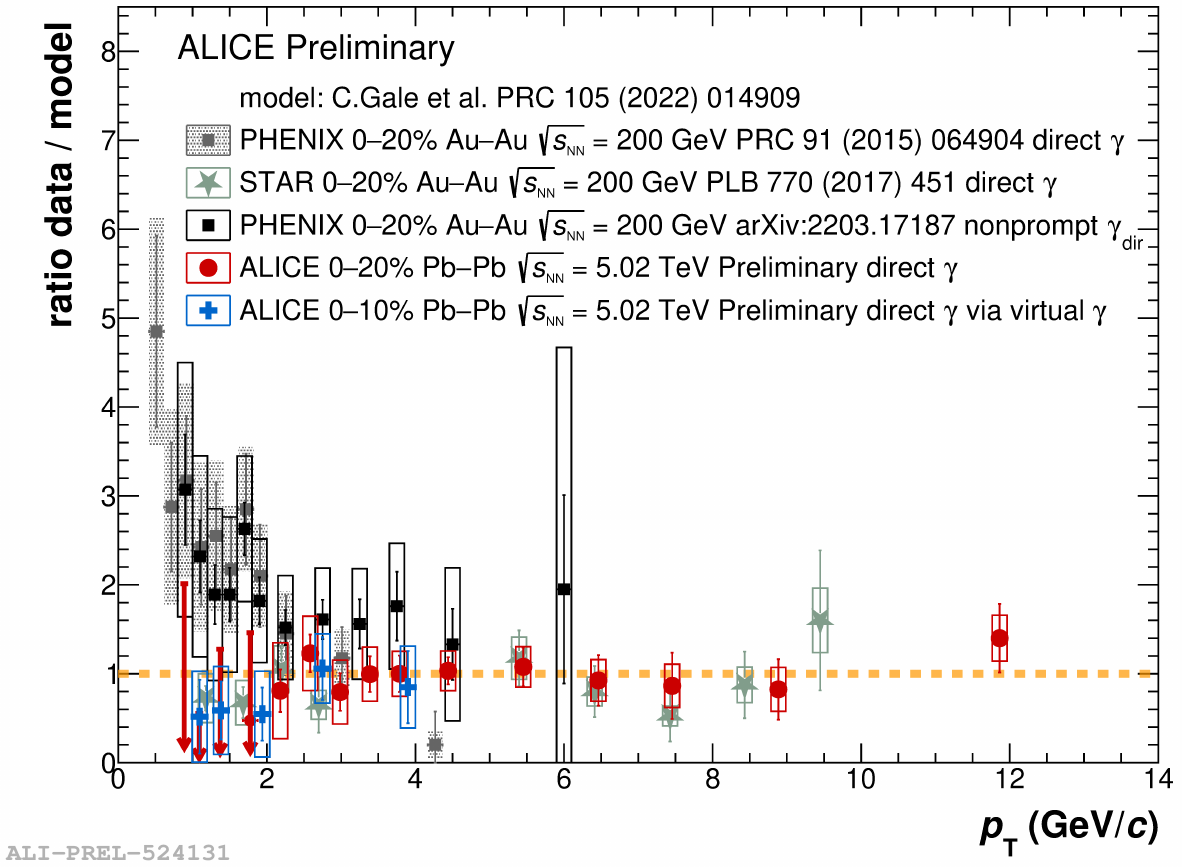}
 		\caption{ The direct photon transverse momentum spectrum measured by the PHENIX, STAR, and ALICE Collaborations, divided by the calculations of \cite{Gale:2021emg}. The figure is from Ref. \cite{Reygers:2022crp} (Creative Common License, CC-BY 4.0).}
 		\label{fig-ALICE}       
 	\end{center}
 \end{figure}

 More recently, the ALICE Collaboration has released data on real direct photons emitted in collisions of Pb + Pb at $\sqrt{s_{\rm NN}}= 5.02$ TeV  analyzed with different methods (photon conversion and virtual photon) \cite{Reygers:2022crp}. Fig. \ref{fig-ALICE} shows those data, together with previous data from the PHENIX and STAR Collaborations at RHIC divided by results of the calculation discussed earlier which also made predictions for the eventual 5.02 TeV measurements \cite{Gale:2021emg}. The new data and the model predictions are in agreement within statistical uncertainties. 
 
 \subsection{Net photon flow}
As is the case for hadrons, the measurement of the anisotropy of photon momentum distributions contains valuable information about the microscopic dynamics of the processes participating in photon production \cite{Chatterjee:2005de}. 
Because of limited statistics,  the measurement and calculation of  photon flow -- or photon momentum anisotropy -- do not proceed via photon-photon correlations (unlike hadrons), but utilize photon-hadron correlations:
\begin{eqnarray}
	v_n (p_{\rm T}^\gamma) = \frac{\langle  v_n^\gamma (p_{\rm T}^\gamma) v_n^h \cos \left(n \left[\Psi_n^\gamma (p_{\rm T}^\gamma) - \Psi_n^h\right]\right)\rangle }{\sqrt{ \langle (v_n^h)^2 \rangle}}
\end{eqnarray}
In the equation above, $v_n^\gamma$ is the photon flow coefficient, $v_n^h$ is that of charged hadrons, and $\Psi_n^i$ is the event plane angle of photons ($i = \gamma$) and hadrons ($i = h$). 
 The hadrons are used to define a reference plane, and the photon momentum anisotropy is measured with respect to this plane \cite{Paquet:2015lta}. Therefore, prior to the consideration of electromagnetic observables and of their momentum anisotropy,  it is verified that the theoretical modeling ot the hadronic sector is in agreement with the corresponding  data. 
 
\begin{figure}[htbp]
 	\begin{center}
 		\includegraphics[width=0.46\textwidth]{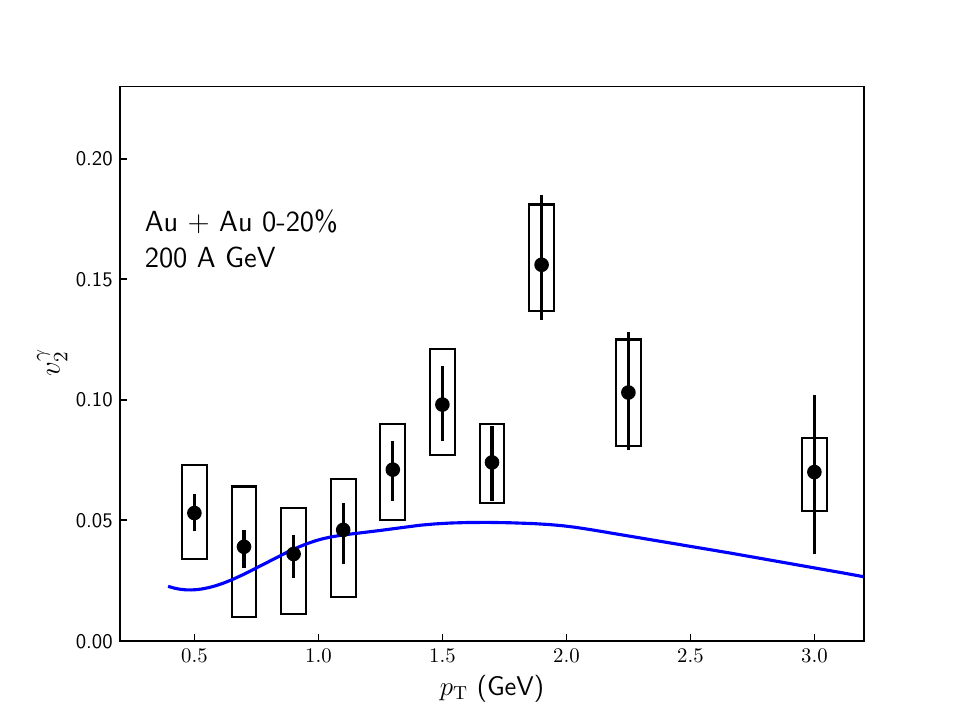} 
 		 		\includegraphics[width=0.46\textwidth]{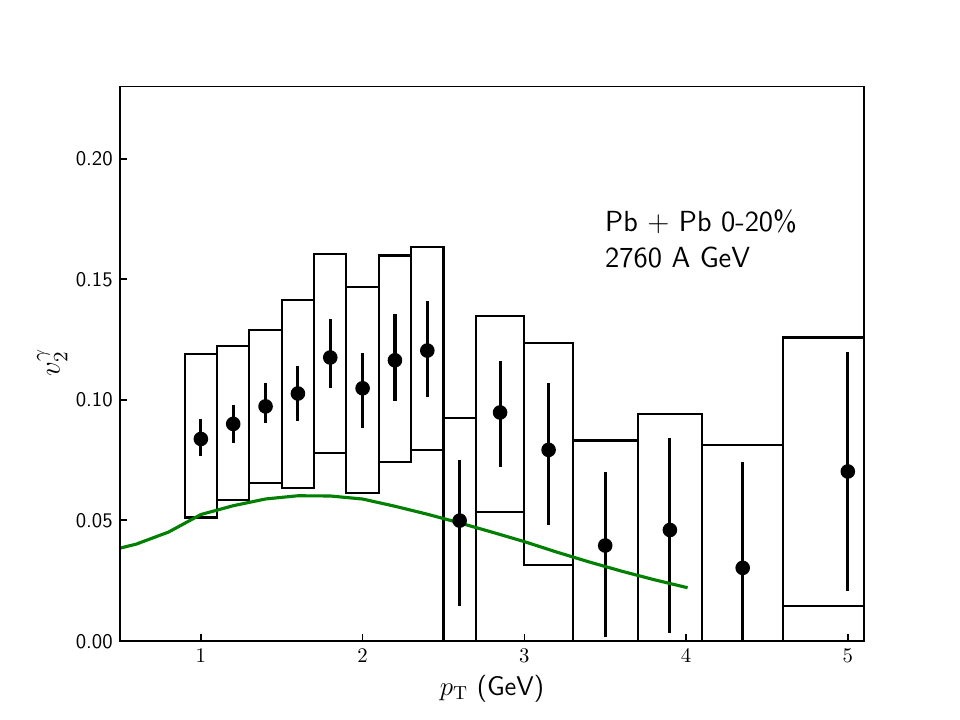} 
 		\caption{(Left panel) The direct photon elliptic flow coefficient, $v_2^\gamma (p_{\rm T})$, for collisions of Au + Au at an energy of 200 $A$ GeV, at midrapidity. The data are measured by the PHENIX Collaboration using the conversion method \cite{PHENIX:2015igl}. The error bars correspond to statistical uncertainties, and the boxes represent systematic uncertainties. (Right panel) The direct photon $v_2^\gamma (p_{\rm T})$ measured by the ALICE Collaboration for  Pb + Pb collisions at 2760 $A$ GeV \cite{ALICE:2018dti}. The error bars are statistical uncertainties and the boxes, the total uncertainty. In both panels the curve is the result of calculations with the approach described in Ref. \cite{Gale:2021emg}, with $\tau_{\rm chem} = 1$ fm/c as shown previously.}
 		\label{fig-photon_v2}       
 	\end{center}
\end{figure}

 Figure \ref{fig-photon_v2} shows the results of calculating photon elliptic flow, $v_2 (p_{\rm T}^\gamma)$, with the multistage model, at RHIC and at LHC energies. The message is that the photon flow measured by the PHENIX Collaboration is underestimated  by the model. The same can be said for the photon flow measured at the LHC, but the statistical significance of the deviation is considerably less. 
A similar difference between theory and data (especially at RHIC) is observed in other calculations using modeling based on either fluid dynamics,  thermal fireballs \cite{vanHees:2014ida}, or PHSD \cite{Linnyk:2015tha}. The tension between data and calculation for the combined photon spectrum and $v_2$ has been termed the ``direct photon puzzle''; a clear solution has yet to be identified  \cite{David:2019wpt}. 

The  direct photon puzzle has driven the development of some new research ideas, while bringing theory and experiments under increased scrutiny \cite{David:2019wpt}. For instance, it has triggered the re-examination of some previous estimates of photon emission around the cross-over temperature from partonic to composite hadronic matter \cite{vanHees:2014ida,Gale:2014dfa,Kim:2016ylr}. Among  the topics put forward to possibly address the puzzle is the effect on electromagnetic radiation of the early strong magnetic fields\footnote{$B \approx 10^{18} G$, the strongest known in Nature  \cite{Skokov:2009qp,Tuchin:2015oka,Huang:2015oca,Adhikari:2024bfa}} generated in non-central heavy ion collisions \cite{McLerran:2013hla,Basar:2012bp,Muller:2013ila,Wang:2020dsr,Ayala:2017vex,Sun:2023rhh}. The cumulative effect of including all known corrections to the photon yield and flow coefficients is still to be worked out, however it is likely that knowing the global effect of strong electromagnetic fields on all observables will likely have to wait for the development of relativistic magneto-hydrodynamics as applied to heavy-ion collisions \cite{Mayer:2024dze,Mayer:2024kkv}. 

\subsection{Dileptons}

 The evolution of thermal rates for dilepton emission \cite{Baier:1988xv,Altherr:1989jc,Aurenche:2002wq,Laine:2013vma} has proceeded in parallel with that of real photons, and the effect of viscosity corrections has been discussed and highlighted there as well \cite{Vujanovic:2016anq,Vujanovic:2017psb,Vujanovic:2019yih}. As was the case for the emission of real photons, not all channels have been corrected for viscous effects. For example, low invariant mass ($M < m_{\phi}$) lepton pairs will receive a large contributions from baryons affecting $\Pi_{\mu \nu}$ in Eq. (\ref{rate_eqs}) \cite{Rapp:2009az}; estimates of viscous effects there are still being performed. In the QGP, the electromagnetic emissivity of an ensemble of hot partons is known today at NLO, and at finite baryon density, $\mu_{\rm B} \ne 0$ \cite{Churchill:2023vpt}. Note that QCD photon rates complete at ${\mathcal O} (\alpha_s)$, $\mu_{\rm B} \ne 0$, and including the LPM effect had  been established  in Ref.  \cite{Gervais:2012wd}. Both of those will be treated in the future for non-equilibrium effects owing to viscosity. 

The dilepton-equivalent of the ``cold photons'' calculable in pQCD, are those originating from the Drell-Yan (DY) process \cite{Drell:1970wh}.  The calculations of DY cross sections have also been integrated in numerical packages such as Pythia \cite{Sjostrand:2006za}, or the more specialized  DYTurbo \cite{Camarda:2019zyx} which performs a transverse-momentum resummation up to next-to-next-to-leading logarithmic accuracy, combined with fixed-order results at next-to-next-to-leading order (NNLO). Those two permit the evaluation of the irreducible background to the thermal dilepton signal.  

Even though the thermal dilepton yield stems from the same current-current correlator as for the real photons, measurements of dileptons offer in addition the potential to reveal the effects of  chiral symmetry restauration \cite{Rapp:1999ej}  as well as the hope to access an in-medium temperature that is impervious to kinematic distortions. The ``dilepton thermometer'' exists because the yield of lepton pairs can be calculated and measured as a function of their invariant mass $M^2~=~(p_+~+~p_-)^\mu~(p_{+}~+~p_{-})_\mu$, whereas real photons  are confined to the light cone and their spectrum depends on their momentum which is not a Lorentz invariant.  For instance, photon spectra will experience Doppler shift: the local motion in the lab frame of a given cell  will distort the fluid rest frame photon distribution. This makes photon and dilepton measurements {\em complementary}: in addition to the effective  temperature information, the photon spectra will also inform the dynamical modeling \cite{vanHees:2011vb,Shen:2013vja,Massen:2024pnj}, while the dilepton spectra will offer a Doppler-free reading. Note however that measurements will still consist of lepton pairs emitted at all stages of Fig. \ref{fig-stages}, and over a distribution of cell temperatures.

In terms of measurements, low and intermediate mass dileptons were previously analyzed by the pioneering DLS experiment \cite{DLS:1997kbk} at the LBNL Bevalac, and at the CERN SPS by the CERES, NA38/NA50, and Helios-3 Collaborations \cite{Specht:2007ez}.  In the low invariant mass region (LMR), $M \lesssim 1$ GeV, the dilepton spectrum measured with great precision by the NA60 Collaboration in collisions of In + In at    $A$ 158 GeV           reveals \cite{NA60:2008ctj} a striking broadening of the $\rho$ meson spectral density, where coupling with baryon resonances  plays a decisive role \cite{Rapp:1999ej}.
At RHIC, dileptons are investigated by the STAR \cite{Geurts:2012rv} and PHENIX \cite{PHENIX:2005kfn} Collaborations, and at the LHC by ALICE \cite{Feuillard:2022ket}. For the collider conditions prevailing  at RHIC, the STAR collaboration has reported dilepton measurements at various energies, namely  BES-I and -II (Beam Energy Scan) and  $\sqrt{s_{\rm NN}}$ =  200 GeV. The LMR measurements agree \cite{CJHP2024} with calculations based on the same spectral density integrated with a thermal fireball model  previously found to match the one extracted by  NA60 from the In + In data \cite{Huck:2014mfa}, also supporting the scenario where $\rho$ meson broadening is driven by interaction with baryon resonances.

The dilepton spectrum in the intermediate mass region (IMR), 1 GeV $<  M \lesssim 3$ GeV has been the focus of some attention because of the possibility for dilepton measurements to reveal a possible restauration of chiral symmetry in a hot and dense strongly interacting system \cite{Rapp:1999ej}, proceeding through  a mixing of the vector and axial-vector spectral densities \cite{Geurts:2022xmk}.  This would  necessitate the observation of the pseudovector spectral density \cite{Kapusta:1993hq,Hohler:2013eba,Geurts:2022xmk}, which has however remained elusive. 
Because a clear signal is still ambiguous, the NA60+ Collaboration has put forward a new experiment proposal which features the search for chiral symmetry restauration effects, the study of the order of the phase transition at large baryochemical potential through the measurement of a caloric curve, and the onset of the deconfinement through the measurement of $J/\psi$  suppression at the CERN SPS. The letter of intent \cite{NA60:2022sze} states that  chiral restauration through $\rho-a_1$ mixing would generate a 20-25\% enhancement of the dilepton invariant mass spectrum over the case where the $\rho$ and the  $a_1$ do not mix in the dilepton invariant mass region around $M \sim $ 1 GeV. If approved, data taking could start in 2029.

\begin{figure}[htbp]
	\begin{center}
		\includegraphics[width=0.6\textwidth]{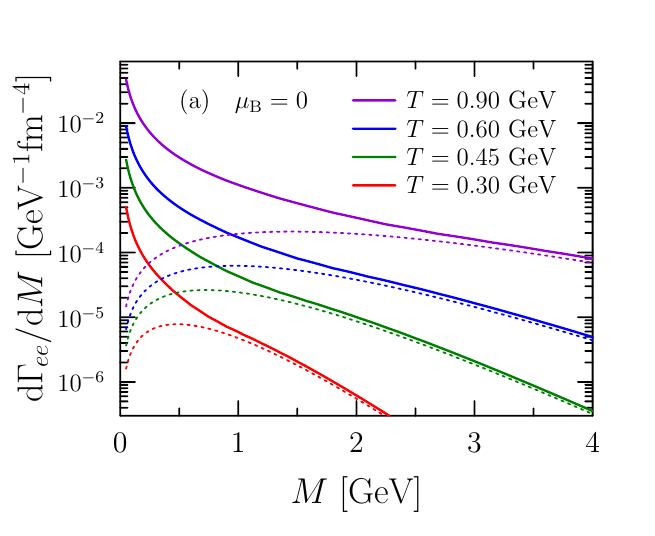}
		\caption{Thermal dilepton emission rates for a static midrapidity source with vanishing net baryon density at different temperatures, as a function of the dilepton invariant mass $M$. The dotted lines represent the emission rate computed at leading order (LO) in the strong coupling (${\cal O} (\alpha_s^0)$), and the solid curves are the emission rates at full NLO (${\cal O} (\alpha_s)$). This figure is adapted from \cite{Churchill:2023vpt}.}
		\label{fig-ratesNLO}       
	\end{center}
\end{figure}

 We leave the observation of chiral symmetry restauration as a tantalizing possibility for now and concentrate on temperature extractions, which has been the subject of previous phenomenological and experimental studies \cite{NA60:2008ctj,Rapp:2014hha,STAR:2024bpc}.  Recent analyses done using the multistage approach introduced earlier have also been completed  \cite{ Churchill:2023zkk,Churchill:2023vpt}. The first lesson from these recent papers is the importance of next-to-leading order (in the strong coupling) QCD contributions to the dilepton spectrum coming from the hot partonic medium, when compared to results obtained at leading order (LO). This is made clear in Fig. \ref{fig-ratesNLO} which shows the thermal lepton pair production rate $\Gamma$ (@LO and @NLO) as a function of invariant mass for different medium temperatures. Clearly, NLO corrections have a considerable effect -- qualitatively and quantitatively, depending on the temperature -- at low ($M < 1$ GeV) invariant mass, but also affect the intermediate mass (1 GeV $\lesssim M \lesssim 3$ GeV) region. Note that the need to go beyond the Born term for dilepton production at finite temperature had been known for some time \cite{Braaten:1989mz,Wong:1991be}. In the future, as was the case for some channels contributing to real photon emission, NLO dilepton emission rates will have to be corrected for non-equilibrium conditions; some of this has been done for LO rates \cite{Vujanovic:2016anq,Vujanovic:2017psb,Kasmaei:2018oag,Vujanovic:2019yih}. 

 The emission of lepton pairs from the pre-hydro era has also received some attention \cite{Martinez:2008di}. Recent estimates based on using either kinetic theory or \kompost (as was done for photons) and including the contribution of the Drell-Yan leptons pairs have been completed \cite{Coquet:2021lca,Garcia-Montero:2024lbl,Wu:2024pba}, and the pre-equilibrium dilepton spectra computed for LHC conditions in those two approaches  have been found to have very similar characteristics \cite{JFHP2024}.  
 The conclusion that follows from those studies is that a pre-equilibrium component can contribute measurably to the spectrum of dileptons in the intermediate mass region at LHC energy (5.02 TeV). Those dileptons will outshine those from the Drell-Yan process; however the unambiguous interpretation  of this signal will greatly benefit from the explicit measurement of lepton pairs coming from the simultaneous decay of open charm mesons, as planned by the next generation of experiments. 
 
 Finally, the dilepton spectrum can be formulated as a double differential, where the $p_T$ spectrum can be studied in various windows of invariant mass, for instance. This additional feature which is absent for real photons will constitute a demand on experimental statistics, but can highlight further dynamical effects, like the chemical equilibrium at early times \cite{Wu:2024pba} or the value of QCD transport parameters \cite{Vujanovic:2016anq}. 

  \begin{figure}[htbp]
	\begin{center}
		\includegraphics[width=0.6\textwidth]{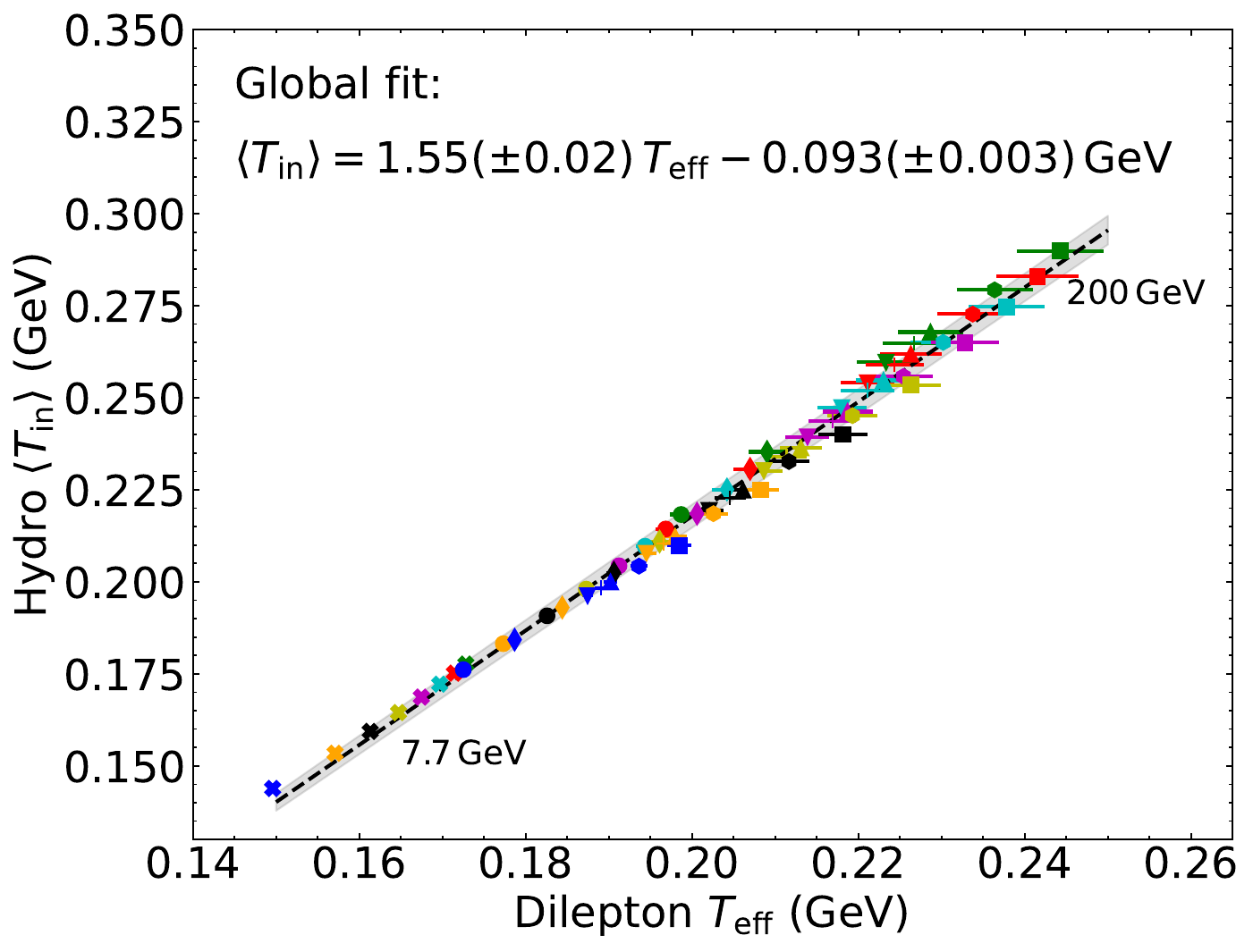}
		\caption{ Correlation between the average initial temperature $\langle T_{\rm in}\rangle$ of the fluid dynamical model  and the effective temperature obtained from the slope of the dilepton spectrum in the intermediate mass region. The plot contains several regions of centrality, and energies from 7.7 to 200 GeV. The black dashed line is a global fit to all data points and the gray band represents the uncertainty related to the fitting procedure. This figure is adapted from Ref. \cite{Churchill:2023zkk}.   }
		\label{fig-Textract}       
	\end{center}
\end{figure}	 

Considering the evolution in the calculation of emission rates and the current sophistication of the multistage modeling approaches, a legitimate question to ask is whether the promise of the early -- possibly optimistic -- estimates \cite{Kajantie:1981wg} of extracting legitimate temperatures from the slope of the dilepton spectrum measured in heavy-ion collisions can be realized, many decades later. A recent study was devoted specifically to this question \cite{Churchill:2023zkk}, where values of temperature extracted from the slope of dilepton spectra in the intermediate mass region  calculated at NLO in QCD  at finite temperature and non-zero baryon chemical potential were compared with values of ``true'' temperature found within the (3+1)D relativistic viscous hydrodynamics simulations. The summary of this study is found in Fig. \ref{fig-Textract}, which shows a clear correlation between the average initial hydrodynamical temperature $\langle T_{\rm in}\rangle$ and $T_{\rm eff}$, the temperature extracted from dilepton spectra produced in Au + Au collisions spanning a range of centralities and energies from 7.7 to 200 GeV. The correlation between ``effective'' and ``true'' temperatures is much tighter than when analyzed with photons \cite{Shen:2013vja}. The figure highlights a linear correlation, even though the average initial temperature $\langle T_{\rm in}\rangle$ is really the central value of some broad distribution \cite{Churchill:2023zkk} over several fluid cells. It is exciting to note that almost all of the extracted values of temperature lie above the pseudo-critical value $T_{\rm pc} \approx 156$ MeV, at $\mu_{\rm B} = 0$ extracted from lattice QCD \cite{Steinbrecher:2018phh}.  The results shown in Fig. \ref{fig-Textract} also drive home the point that even something as physical as temperature will need a realistic theoretical model tuned to other data (hadronic and otherwise) to be interpreted properly. 

Finally, the flow coefficients ($v_n (p_{\rm T})$) of dileptons also contain previous information, in particular since their spectrum is in principle a function of invariant mass and transverse momentum \cite{Chatterjee:2007xk,Kasmaei:2018oag,Vujanovic:2016anq,Vujanovic:2017psb,Vujanovic:2019yih}. Preliminary flow measurements  have been reported \cite{Geurts:2013kiz} by the STAR Collaboration at RHIC. 

Another aspect of electromagnetic signals from heavy-ion collisions  that is bound to receive more attention is that of polarization.  For heavy-ion collisions at lower energies that those studied at RHIC and at the LHC -- those at SIS and at the CERN SPS -- the angular lepton distributions reflecting the virtual photon polarization have been studied \cite{Speranza:2017hso,Seck:2023oyt} using coarse-grained transport models, and have been  measured by the NA60 \cite{NA60:2008iqj} and HADES \cite{Seck:2023oyt} Collaborations. At those energies, the angular distribution of low invariant-mass dileptons is consistent with the presence of the resonances generated by the interactions of the composite hadrons, with little or no apparent sign of a QGP signal. In collider conditions, the polarization of the virtual photons produced in quark-antiquark annihilation at leading order has been studied in theory and used as a probe of  the QGP momentum anisotropy  \cite{Baym:2017qxy,Coquet:2023wjk}, and so has that of real photons \cite{Hauksson:2023dwh}. 

However, as was the case for the dilepton mass spectra, polarization studies of the QGP demand NLO corrections in the low and intermediate invariant mass regions: as $M \to 0$, the LO Born term $q \bar{q} \to \gamma^*$ has no phase space to produce a real photon. This is illustrated in the following: the virtual photon polarization can be observed via the angular distribution of final-state leptons in a given Lorentz frame. In the rest frame of the virtual photon, the helicity frame, the coefficient of the polar angle distribution of one of the leptons in the dilepton is $\lambda_\theta$ and can be shown to be related\footnote{In a static medium, for lepton and antilepton with negligible rest mass.} to the transverse (T) and longitudinal (L) components of the dilepton spectral density \cite{Wu:2024vyc}: $\lambda_\theta \sim \left(\rho_{\rm T} - \rho_{\rm L}\right)/ \left(\rho_{\rm T} + \rho_{\rm L}\right)$. There again, one observes stark qualitative and quantitative differences between $\lambda_\theta^{\rm LO}$ and $\lambda_\theta^{\rm NLO}$. One can show that $\lim_{M \to 0} \lambda_\theta^{\rm NLO} = 1$, because $\lim_{M \to 0} \rho_{\rm L}^{\rm NLO} = 0$, as expected of a real photon. On the other hand,  $\lim_{M \to 0} \lambda_\theta^{\rm LO} \neq 1 $  \cite{Wu:2024vyc}. Finally, it is encouraging to note that, after properly averaging  $\lambda_\theta$ over hydrodynamical fluid cells using a realistic space-time model, the  characteristic profile survives the fluid evolution \cite{Wu:2024vyc}: clearly, electromagnetic polarization is another  precious variable to signal the presence of, and to characterize, the QGP.

\section{Conclusions}

It is evident that during the last five decades, the progress in theory and  measurement of electromagnetic radiation emitted in relativistic heavy-ion collisions has been striking. This  owes much to  advances in the treatment of electromagnetic emissivity, but this progress was realized concurrently with rapid developments in the modeling of relativistic nuclear collisions. It is exciting that data has suggested the existence of a clear thermal signal \cite{NA60:2007lzy} which is on the verge of precise quantification \cite{Churchill:2023vpt}.     Photons and dileptons can act as tomographic probes which can also inform our understanding of the space-time evolution of the interaction  volume. Given the many sources of electromagnetic quanta and the sophistication of modern multistage modeling of relativistic heavy-ion collisions, the fact that photon spectrum calculations and measurements show some level of agreement -- even if  improvement is warranted  -- is nothing short of remarkable. This is indeed the era of multimessenger heavy-ion physics. What remains to be done in theory is  integrating the many stages we have described in this paper into an end-to-end approach. This is not a small task but the roadmap is clear and if recent developments are any indication, its realization is within reach. While experimental measurements of electromagnetic radiation are complicated because of small signal strength and large backgrounds, the community has spent significant effort to systematically  include them alongside hadronic observables.

There are several aspects of the emission of electromagnetic radiation in heavy-ion collisions that this article could not cover. A few examples are that of electromagnetic interferometry \cite{KumarSrivastava:1993mp,Garcia-Montero:2019kjk}, of real and virtual photon emission from small systems \cite{Shen:2016zpp},  of radiation from from jet-medium interactions \cite{Srivastava:2002ic,Yazdi:2022cuk}, and the radiation of photons with very low momenta \cite{Bailhache:2024mck}.  Clearly, the study of electromagnetic emission from nuclear collisions is a rich and still evolving topic which is perhaps only now entering its golden age.

\begin{acknowledgement}
I am happy to acknowledge  many collaborators over the years spent on this topic.  I am also grateful for valuable comments from Han Gao, Greg Jackson, Jean-Fran\c{c}ois Paquet, Bj\"orn Schenke, and Xiang-Yu Wu. 
This work was funded in part by the Natural Sciences and Engineering Research Council of Canada (NSERC) [SAPIN-2020-00048]. 
\end{acknowledgement}
%

\input{references}
\end{document}

%% file: references.tex
%
%
 \bibliographystyle{spphys}
\bibliography{refs}
%

%
%



%


%% file: EM_probes2.bbl
\begin{thebibliography}{100}
\providecommand{\url}[1]{{#1}}
\providecommand{\urlprefix}{URL }
\expandafter\ifx\csname urlstyle\endcsname\relax
  \providecommand{\doi}[1]{DOI \discretionary{}{}{}#1}\else
  \providecommand{\doi}{DOI \discretionary{}{}{}\begingroup
  \urlstyle{rm}\Url}\fi

\bibitem{Field:1989uq}
R.D. Field, \emph{Applications of perturbative QCD}.
\newblock Frontiers in physics ; v. 77 (Addison-Wesley, Reading, Mass, 1989)

\bibitem{Drell:1970wh}
S.D. Drell, T.M. Yan, Phys. Rev. Lett. \textbf{25}, 316 (1970).
\newblock \doi{10.1103/PhysRevLett.25.316}.
\newblock [Erratum: Phys.Rev.Lett. 25, 902 (1970)]

\bibitem{Owens:1986mp}
J.F. Owens, Rev. Mod. Phys. \textbf{59}, 465 (1987).
\newblock \doi{10.1103/RevModPhys.59.465}

\bibitem{Schenke:2010nt}
B.~Schenke, S.~Jeon, C.~Gale, Phys. Rev. C \textbf{82}, 014903 (2010).
\newblock \doi{10.1103/PhysRevC.82.014903}

\bibitem{Bleicher:1999xi}
M.~Bleicher, et~al., J. Phys. G \textbf{25}, 1859 (1999).
\newblock \doi{10.1088/0954-3899/25/9/308}

\bibitem{Shuryak:1978ij}
E.V. Shuryak, Phys. Lett. B \textbf{78}, 150 (1978).
\newblock \doi{10.1016/0370-2693(78)90370-2}

\bibitem{Kajantie:1981wg}
K.~Kajantie, H.I. Miettinen, Z. Phys. C \textbf{9}, 341 (1981).
\newblock \doi{10.1007/BF01548770}

\bibitem{Domokos:1980ba}
G.~Domokos, J.I. Goldman, Phys. Rev. D \textbf{23}, 203 (1981).
\newblock \doi{10.1103/PhysRevD.23.203}

\bibitem{Kajantie:1986dh}
K.~Kajantie, J.I. Kapusta, L.D. McLerran, A.~Mekjian, Phys. Rev. D \textbf{34},
  2746 (1986).
\newblock \doi{10.1103/PhysRevD.34.2746}

\bibitem{Feinberg:1976ua}
E.L. Feinberg, Nuovo Cim. A \textbf{34}, 391 (1976)

\bibitem{McLerran:1984ay}
L.D. McLerran, T.~Toimela, Phys. Rev. D \textbf{31}, 545 (1985).
\newblock \doi{10.1103/PhysRevD.31.545}

\bibitem{Weldon:1990iw}
H.A. Weldon, Phys. Rev. D \textbf{42}, 2384 (1990).
\newblock \doi{10.1103/PhysRevD.42.2384}

\bibitem{Gale:1990pn}
C.~Gale, J.I. Kapusta, Nucl. Phys. B \textbf{357}, 65 (1991).
\newblock \doi{10.1016/0550-3213(91)90459-B}

\bibitem{Kapusta:2007xjq}
J.I. Kapusta, C.~Gale, \emph{{Finite-Temperature Field Theory : Principles and
  Applications, 2nd edition}} (Cambridge University Press, 2007).
\newblock \doi{10.1017/9781009401968}

\bibitem{Ghiglieri:2016tvj}
J.~Ghiglieri, O.~Kaczmarek, M.~Laine, F.~Meyer, Phys. Rev. D \textbf{94}(1),
  016005 (2016).
\newblock \doi{10.1103/PhysRevD.94.016005}

\bibitem{Ce:2022fot}
M.~C\`e, T.~Harris, A.~Krasniqi, H.B. Meyer, C.~T\"or\"ok, Phys. Rev. D
  \textbf{106}(5), 054501 (2022).
\newblock \doi{10.1103/PhysRevD.106.054501}

\bibitem{Ali:2024xae}
S.~Ali, D.~Bala, A.~Francis, G.~Jackson, O.~Kaczmarek, J.~Turnwald, T.~Ueding,
  N.~Wink, Phys. Rev. D \textbf{110}(5), 054518 (2024).
\newblock \doi{10.1103/PhysRevD.110.054518}

\bibitem{Weldon:1983jn}
H.A. Weldon, Phys. Rev. D \textbf{28}, 2007 (1983).
\newblock \doi{10.1103/PhysRevD.28.2007}

\bibitem{Majumder:2001iy}
A.~Majumder, C.~Gale, Phys. Rev. C \textbf{65}, 055203 (2002).
\newblock \doi{10.1103/PhysRevC.65.055203}

\bibitem{Wong:2000hq}
S.M.H. Wong, Phys. Rev. D \textbf{64}, 025007 (2001).
\newblock \doi{10.1103/PhysRevD.64.025007}

\bibitem{Sinha:1983jm}
B.~Sinha, Phys. Lett. B \textbf{128}, 91 (1983).
\newblock \doi{10.1016/0370-2693(83)90080-1}

\bibitem{Hwa:1985xg}
R.C. Hwa, K.~Kajantie, Phys. Rev. D \textbf{32}, 1109 (1985).
\newblock \doi{10.1103/PhysRevD.32.1109}

\bibitem{Staadt:1985uc}
G.~Staadt, W.~Greiner, J.~Rafelski, Phys. Rev. D \textbf{33}, 66 (1986).
\newblock \doi{10.1103/PhysRevD.33.66}

\bibitem{Kapusta:1991qp}
J.I. Kapusta, P.~Lichard, D.~Seibert, Phys. Rev. D \textbf{44}, 2774 (1991).
\newblock \doi{10.1103/PhysRevD.47.4171}.
\newblock [Erratum: Phys.Rev.D 47, 4171 (1993)]

\bibitem{Baier:1991em}
R.~Baier, H.~Nakkagawa, A.~Niegawa, K.~Redlich, Z. Phys. C \textbf{53}, 433
  (1992).
\newblock \doi{10.1007/BF01625902}

\bibitem{Braaten:1989mz}
E.~Braaten, R.D. Pisarski, Nucl. Phys. B \textbf{337}, 569 (1990).
\newblock \doi{10.1016/0550-3213(90)90508-B}

\bibitem{Aurenche:1998nw}
P.~Aurenche, F.~Gelis, R.~Kobes, H.~Zaraket, Phys. Rev. D \textbf{58}, 085003
  (1998).
\newblock \doi{10.1103/PhysRevD.58.085003}

\bibitem{Arnold:2001ms}
P.B. Arnold, G.D. Moore, L.G. Yaffe, JHEP \textbf{12}, 009 (2001).
\newblock \doi{10.1088/1126-6708/2001/12/009}

\bibitem{Ghiglieri:2013gia}
J.~Ghiglieri, J.~Hong, A.~Kurkela, E.~Lu, G.D. Moore, D.~Teaney, JHEP
  \textbf{05}, 010 (2013).
\newblock \doi{10.1007/JHEP05(2013)010}

\bibitem{Meissner:1987ge}
U.G. Meissner, Phys. Rept. \textbf{161}, 213 (1988).
\newblock \doi{10.1016/0370-1573(88)90090-7}

\bibitem{Song:1993ae}
C.~Song, Phys. Rev. C \textbf{47}, 2861 (1993).
\newblock \doi{10.1103/PhysRevC.47.2861}

\bibitem{Turbide:2003si}
S.~Turbide, R.~Rapp, C.~Gale, Phys. Rev. C \textbf{69}, 014903 (2004).
\newblock \doi{10.1103/PhysRevC.69.014903}

\bibitem{Heffernan:2014mla}
M.~Heffernan, P.~Hohler, R.~Rapp, Phys. Rev. C \textbf{91}(2), 027902 (2015).
\newblock \doi{10.1103/PhysRevC.91.027902}

\bibitem{Holt:2015cda}
N.P.M. Holt, P.M. Hohler, R.~Rapp, Nucl. Phys. A \textbf{945}, 1 (2016).
\newblock \doi{10.1016/j.nuclphysa.2015.09.008}

\bibitem{Rapp:1999ej}
R.~Rapp, J.~Wambach, Adv. Nucl. Phys. \textbf{25}, 1 (2000)

\bibitem{Michler:2012mg}
F.~Michler, H.~van Hees, D.D. Dietrich, S.~Leupold, C.~Greiner, Annals Phys.
  \textbf{336}, 331 (2013).
\newblock \doi{10.1016/j.aop.2013.05.021}

\bibitem{Schenke:2005ry}
B.~Schenke, C.~Greiner, Phys. Rev. C \textbf{73}, 034909 (2006).
\newblock \doi{10.1103/PhysRevC.73.034909}

\bibitem{Aurenche:2006vj}
P.~Aurenche, M.~Fontannaz, J.P. Guillet, E.~Pilon, M.~Werlen, Phys. Rev. D
  \textbf{73}, 094007 (2006).
\newblock \doi{10.1103/PhysRevD.73.094007}

\bibitem{Arleo:2011gc}
F.~Arleo, K.J. Eskola, H.~Paukkunen, C.A. Salgado, JHEP \textbf{04}, 055
  (2011).
\newblock \doi{10.1007/JHEP04(2011)055}

\bibitem{Klasen:2013mga}
M.~Klasen, C.~Klein-B\"osing, F.~K\"onig, J.P. Wessels, JHEP \textbf{10}, 119
  (2013).
\newblock \doi{10.1007/JHEP10(2013)119}

\bibitem{Paakkinen:2017jpo}
P.~Paakkinen, Frascati Phys. Ser. pp. 33--40 (2017)

\bibitem{Klasen:2014xfa}
M.~Klasen, F.~K\"onig, Eur. Phys. J. C \textbf{74}(8), 3009 (2014).
\newblock \doi{10.1140/epjc/s10052-014-3009-x}

\bibitem{Paquet:2015lta}
J.F. Paquet, C.~Shen, G.S. Denicol, M.~Luzum, B.~Schenke, S.~Jeon, C.~Gale,
  Phys. Rev. C \textbf{93}(4), 044906 (2016).
\newblock \doi{10.1103/PhysRevC.93.044906}

\bibitem{Gale:2013da}
C.~Gale, S.~Jeon, B.~Schenke, Int. J. Mod. Phys. A \textbf{28}, 1340011 (2013).
\newblock \doi{10.1142/S0217751X13400113}

\bibitem{Denicol}
G.S. Denicol, D.H. Rischke, \emph{Microscopic foundations of relativistic fluid
  dynamics} (Springer, 2022)

\bibitem{Chapman}
S.~Chapman, T.G. Cowling, \emph{The mathematical theory of non-uniform gases:
  an account of the kinetic theory of viscosity, thermal conduction and
  diffusion in gases} (Cambridge university press, 1990)

\bibitem{Grad}
H.~Grad, Communications on pure and applied mathematics \textbf{2}(4), 331
  (1949)

\bibitem{Dion:2011pp}
M.~Dion, J.F. Paquet, B.~Schenke, C.~Young, S.~Jeon, C.~Gale, Phys. Rev. C
  \textbf{84}, 064901 (2011).
\newblock \doi{10.1103/PhysRevC.84.064901}

\bibitem{HeinzSchenke}
U. Heinz and B. Schenke, this volume

\bibitem{Shen:2014nfa}
C.~Shen, J.F. Paquet, U.~Heinz, C.~Gale, Phys. Rev. C \textbf{91}(1), 014908
  (2015).
\newblock \doi{10.1103/PhysRevC.91.014908}

\bibitem{Schwinger1961}
J.~Schwinger, Journal of Mathematical Physics \textbf{2}(3), 407 (1961).
\newblock \doi{10.1063/1.1703727}.
\newblock \urlprefix\url{https://doi.org/10.1063/1.1703727}

\bibitem{Keldysh:1964ud}
L.V. Keldysh, Zh. Eksp. Teor. Fiz. \textbf{47}, 1515 (1964)

\bibitem{Serreau:2003wr}
J.~Serreau, JHEP \textbf{05}, 078 (2004).
\newblock \doi{10.1088/1126-6708/2004/05/078}

\bibitem{Hauksson:2017udm}
S.~Hauksson, S.~Jeon, C.~Gale, Phys. Rev. C \textbf{97}(1), 014901 (2018).
\newblock \doi{10.1103/PhysRevC.97.014901}

\bibitem{Hauksson:2020wsm}
S.~Hauksson, S.~Jeon, C.~Gale, Phys. Rev. C \textbf{103}, 064904 (2021).
\newblock \doi{10.1103/PhysRevC.103.064904}

\bibitem{Chiu:2012ij}
M.~Chiu, T.K. Hemmick, V.~Khachatryan, A.~Leonidov, J.~Liao, L.~McLerran, Nucl.
  Phys. A \textbf{900}, 16 (2013).
\newblock \doi{10.1016/j.nuclphysa.2013.01.014}

\bibitem{Linnyk:2015tha}
O.~Linnyk, V.~Konchakovski, T.~Steinert, W.~Cassing, E.L. Bratkovskaya, Phys.
  Rev. C \textbf{92}(5), 054914 (2015).
\newblock \doi{10.1103/PhysRevC.92.054914}

\bibitem{Greif:2016jeb}
M.~Greif, F.~Senzel, H.~Kremer, K.~Zhou, C.~Greiner, Z.~Xu, Phys. Rev. C
  \textbf{95}(5), 054903 (2017).
\newblock \doi{10.1103/PhysRevC.95.054903}

\bibitem{Vovchenko:2016mtf}
V.~Vovchenko, L.G. Pang, H.~Niemi, I.A. Karpenko, M.I. Gorenstein, L.M.
  Satarov, I.N. Mishustin, B.~K\"ampfer, H.~Stoecker, PoS \textbf{BORMIO2016},
  039 (2016).
\newblock \doi{10.22323/1.272.0039}

\bibitem{Berges:2017fsa}
J.~Berges, K.~Reygers, N.~Tanji, R.~Venugopalan, Nucl. Phys. A \textbf{967},
  708 (2017).
\newblock \doi{10.1016/j.nuclphysa.2017.04.034}

\bibitem{Oliva:2017pri}
L.~Oliva, M.~Ruggieri, S.~Plumari, F.~Scardina, G.X. Peng, V.~Greco, Phys. Rev.
  C \textbf{96}(1), 014914 (2017).
\newblock \doi{10.1103/PhysRevC.96.014914}

\bibitem{Garcia-Montero:2023lrd}
O.~Garcia-Montero, A.~Mazeliauskas, P.~Plaschke, S.~Schlichting, JHEP
  \textbf{03}, 053 (2024).
\newblock \doi{10.1007/JHEP03(2024)053}

\bibitem{Cleymans:1992kb}
J.~Cleymans, V.V. Goloviznin, K.~Redlich, Phys. Rev. D \textbf{47}, 173 (1993).
\newblock \doi{10.1103/PhysRevD.47.173}

\bibitem{Kurkela:2018vqr}
A.~Kurkela, A.~Mazeliauskas, J.F. Paquet, S.~Schlichting, D.~Teaney, Phys. Rev.
  C \textbf{99}(3), 034910 (2019).
\newblock \doi{10.1103/PhysRevC.99.034910}

\bibitem{Kurkela:2018wud}
A.~Kurkela, A.~Mazeliauskas, J.F. Paquet, S.~Schlichting, D.~Teaney, Phys. Rev.
  Lett. \textbf{122}(12), 122302 (2019).
\newblock \doi{10.1103/PhysRevLett.122.122302}

\bibitem{Arnold:2002zm}
P.B. Arnold, G.D. Moore, L.G. Yaffe, JHEP \textbf{01}, 030 (2003).
\newblock \doi{10.1088/1126-6708/2003/01/030}

\bibitem{Gale:2021emg}
C.~Gale, J.F. Paquet, B.~Schenke, C.~Shen, Phys. Rev. C \textbf{105}(1), 014909
  (2022).
\newblock \doi{10.1103/PhysRevC.105.014909}

\bibitem{Bazavov_2014}
A.~Bazavov, T.~Bhattacharya, C.~DeTar, H.T. Ding, S.~Gottlieb, R.~Gupta,
  P.~Hegde, U.~Heller, F.~Karsch, E.~Laermann, L.~Levkova, S.~Mukherjee,
  P.~Petreczky, C.~Schmidt, C.~Schroeder, R.~Soltz, W.~Soeldner, R.~Sugar,
  M.~Wagner, P.~Vranas, Physical Review D \textbf{90}(9) (2014).
\newblock \doi{10.1103/physrevd.90.094503}.
\newblock \urlprefix\url{http://dx.doi.org/10.1103/PhysRevD.90.094503}

\bibitem{Schenke:2012wb}
B.~Schenke, P.~Tribedy, R.~Venugopalan, Phys. Rev. Lett. \textbf{108}, 252301
  (2012).
\newblock \doi{10.1103/PhysRevLett.108.252301}

\bibitem{JETSCAPE:2020shq}
D.~Everett, et~al., Phys. Rev. Lett. \textbf{126}(24), 242301 (2021).
\newblock \doi{10.1103/PhysRevLett.126.242301}

\bibitem{Nijs:2020roc}
G.~Nijs, W.~van~der Schee, U.~G\"ursoy, R.~Snellings, Phys. Rev. C
  \textbf{103}(5), 054909 (2021).
\newblock \doi{10.1103/PhysRevC.103.054909}

\bibitem{Heffernan:2023gye}
M.R. Heffernan, C.~Gale, S.~Jeon, J.F. Paquet, Phys. Rev. Lett.
  \textbf{132}(25), 252301 (2024).
\newblock \doi{10.1103/PhysRevLett.132.252301}

\bibitem{WA93:1998uhf}
M.M. Aggarwal, et~al., Phys. Rev. C \textbf{58}, 1146 (1998).
\newblock \doi{10.1103/PhysRevC.58.1146}

\bibitem{WA98:1999rbo}
M.M. Aggarwal, et~al., Phys. Lett. B \textbf{458}, 422 (1999).
\newblock \doi{10.1016/S0370-2693(99)00560-2}

\bibitem{Johnson:2002xj}
I.J. Johnson, Nucl. Phys. A \textbf{715}, 691 (2003).
\newblock \doi{10.1016/S0375-9474(02)01468-9}

\bibitem{Reygers:2002kc}
K.~Reygers, Nucl. Phys. A \textbf{715}, 683 (2003).
\newblock \doi{10.1016/S0375-9474(02)01466-5}

\bibitem{Wilde:2012wc}
M.~Wilde, Nucl. Phys. A \textbf{904-905}, 573c (2013).
\newblock \doi{10.1016/j.nuclphysa.2013.02.079}

\bibitem{Gotz:2021dco}
N.~G\"otz, A.~Sch\"afer, O.~Garcia-Montero, J.F. Paquet, H.~Elfner, C.~Gale,
  Phys. Rev. C \textbf{105}(4), 044910 (2022).
\newblock \doi{10.1103/PhysRevC.105.044910}.
\newblock [Erratum: Phys.Rev.C 109, 049901 (2024)]

\bibitem{PHENIX:2014nkk}
A.~Adare, et~al., Phys. Rev. C \textbf{91}(6), 064904 (2015).
\newblock \doi{10.1103/PhysRevC.91.064904}

\bibitem{STAR:2016use}
L.~Adamczyk, et~al., Phys. Lett. B \textbf{770}, 451 (2017).
\newblock \doi{10.1016/j.physletb.2017.04.050}

\bibitem{ALICE:2015xmh}
J.~Adam, et~al., Phys. Lett. B \textbf{754}, 235 (2016).
\newblock \doi{10.1016/j.physletb.2016.01.020}

\bibitem{David:2019wpt}
G.~David, Rept. Prog. Phys. \textbf{83}(4), 046301 (2020).
\newblock \doi{10.1088/1361-6633/ab6f57}

\bibitem{Reygers:2022crp}
K.~Reygers, Acta Phys. Polon. Supp. \textbf{16}(1), 1 (2023).
\newblock \doi{10.5506/APhysPolBSupp.16.1-A19}

\bibitem{Chatterjee:2005de}
R.~Chatterjee, E.S. Frodermann, U.W. Heinz, D.K. Srivastava, Phys. Rev. Lett.
  \textbf{96}, 202302 (2006).
\newblock \doi{10.1103/PhysRevLett.96.202302}

\bibitem{PHENIX:2015igl}
A.~Adare, et~al., Phys. Rev. C \textbf{94}(6), 064901 (2016).
\newblock \doi{10.1103/PhysRevC.94.064901}

\bibitem{ALICE:2018dti}
S.~Acharya, et~al., Phys. Lett. B \textbf{789}, 308 (2019).
\newblock \doi{10.1016/j.physletb.2018.11.039}

\bibitem{vanHees:2014ida}
H.~van Hees, M.~He, R.~Rapp, Nucl. Phys. A \textbf{933}, 256 (2015).
\newblock \doi{10.1016/j.nuclphysa.2014.09.009}

\bibitem{Gale:2014dfa}
C.~Gale, Y.~Hidaka, S.~Jeon, S.~Lin, J.F. Paquet, R.D. Pisarski, D.~Satow, V.V.
  Skokov, G.~Vujanovic, Phys. Rev. Lett. \textbf{114}, 072301 (2015).
\newblock \doi{10.1103/PhysRevLett.114.072301}

\bibitem{Kim:2016ylr}
Y.M. Kim, C.H. Lee, D.~Teaney, I.~Zahed, Phys. Rev. C \textbf{96}(1), 015201
  (2017).
\newblock \doi{10.1103/PhysRevC.96.015201}

\bibitem{Skokov:2009qp}
V.~Skokov, A.Y. Illarionov, V.~Toneev, Int. J. Mod. Phys. A \textbf{24}, 5925
  (2009).
\newblock \doi{10.1142/S0217751X09047570}

\bibitem{Tuchin:2015oka}
K.~Tuchin, Phys. Rev. C \textbf{93}(1), 014905 (2016).
\newblock \doi{10.1103/PhysRevC.93.014905}

\bibitem{Huang:2015oca}
X.G. Huang, Rept. Prog. Phys. \textbf{79}(7), 076302 (2016).
\newblock \doi{10.1088/0034-4885/79/7/076302}

\bibitem{Adhikari:2024bfa}
P.~Adhikari, et~al.,   (2024)

\bibitem{McLerran:2013hla}
L.~McLerran, V.~Skokov, Nucl. Phys. A \textbf{929}, 184 (2014).
\newblock \doi{10.1016/j.nuclphysa.2014.05.008}

\bibitem{Basar:2012bp}
G.~Basar, D.~Kharzeev, D.~Kharzeev, V.~Skokov, Phys. Rev. Lett. \textbf{109},
  202303 (2012).
\newblock \doi{10.1103/PhysRevLett.109.202303}

\bibitem{Muller:2013ila}
B.~Muller, S.Y. Wu, D.L. Yang, Phys. Rev. D \textbf{89}(2), 026013 (2014).
\newblock \doi{10.1103/PhysRevD.89.026013}

\bibitem{Wang:2020dsr}
X.~Wang, I.A. Shovkovy, L.~Yu, M.~Huang, Phys. Rev. D \textbf{102}(7), 076010
  (2020).
\newblock \doi{10.1103/PhysRevD.102.076010}

\bibitem{Ayala:2017vex}
A.~Ayala, J.D. Castano-Yepes, C.A. Dominguez, L.A. Hernandez,
  S.~Hernandez-Ortiz, M.E. Tejeda-Yeomans, Phys. Rev. D \textbf{96}(1), 014023
  (2017).
\newblock \doi{10.1103/PhysRevD.96.014023}.
\newblock [Erratum: Phys.Rev.D 96, 119901 (2017)]

\bibitem{Sun:2023rhh}
J.A. Sun, L.~Yan, Phys. Rev. C \textbf{109}(3), 034917 (2024).
\newblock \doi{10.1103/PhysRevC.109.034917}

\bibitem{Mayer:2024dze}
M.~Mayer, A.~Dash, G.~Inghirami, H.~Elfner, L.~Rezzolla, D.H. Rischke,   (2024)

\bibitem{Mayer:2024kkv}
M.~Mayer, A.~Dash, G.~Inghirami, H.~Elfner, L.~Rezzolla, D.H. Rischke,   (2024)

\bibitem{Baier:1988xv}
R.~Baier, B.~Pire, D.~Schiff, Phys. Rev. D \textbf{38}, 2814 (1988).
\newblock \doi{10.1103/PhysRevD.38.2814}

\bibitem{Altherr:1989jc}
T.~Altherr, P.~Aurenche, Z. Phys. C \textbf{45}, 99 (1989).
\newblock \doi{10.1007/BF01556676}

\bibitem{Aurenche:2002wq}
P.~Aurenche, F.~Gelis, G.D. Moore, H.~Zaraket, JHEP \textbf{12}, 006 (2002).
\newblock \doi{10.1088/1126-6708/2002/12/006}

\bibitem{Laine:2013vma}
M.~Laine, JHEP \textbf{11}, 120 (2013).
\newblock \doi{10.1007/JHEP11(2013)120}

\bibitem{Vujanovic:2016anq}
G.~Vujanovic, J.F. Paquet, G.S. Denicol, M.~Luzum, S.~Jeon, C.~Gale, Phys. Rev.
  C \textbf{94}(1), 014904 (2016).
\newblock \doi{10.1103/PhysRevC.94.014904}

\bibitem{Vujanovic:2017psb}
G.~Vujanovic, G.S. Denicol, M.~Luzum, S.~Jeon, C.~Gale, Phys. Rev. C
  \textbf{98}(1), 014902 (2018).
\newblock \doi{10.1103/PhysRevC.98.014902}

\bibitem{Vujanovic:2019yih}
G.~Vujanovic, J.F. Paquet, C.~Shen, G.S. Denicol, S.~Jeon, C.~Gale, U.~Heinz,
  Phys. Rev. C \textbf{101}, 044904 (2020).
\newblock \doi{10.1103/PhysRevC.101.044904}

\bibitem{Rapp:2009az}
R.~Rapp, PoS \textbf{CPOD2009}, 040 (2009).
\newblock \doi{10.22323/1.071.0040}

\bibitem{Churchill:2023vpt}
J.~Churchill, L.~Du, C.~Gale, G.~Jackson, S.~Jeon, Phys. Rev. C
  \textbf{109}(4), 044915 (2024).
\newblock \doi{10.1103/PhysRevC.109.044915}

\bibitem{Gervais:2012wd}
H.~Gervais, S.~Jeon, Phys. Rev. C \textbf{86}, 034904 (2012).
\newblock \doi{10.1103/PhysRevC.86.034904}

\bibitem{Sjostrand:2006za}
T.~Sjostrand, S.~Mrenna, P.Z. Skands, JHEP \textbf{05}, 026 (2006).
\newblock \doi{10.1088/1126-6708/2006/05/026}

\bibitem{Camarda:2019zyx}
S.~Camarda, et~al., Eur. Phys. J. C \textbf{80}(3), 251 (2020).
\newblock \doi{10.1140/epjc/s10052-020-7757-5}.
\newblock [Erratum: Eur.Phys.J.C 80, 440 (2020)]

\bibitem{vanHees:2011vb}
H.~van Hees, C.~Gale, R.~Rapp, Phys. Rev. C \textbf{84}, 054906 (2011).
\newblock \doi{10.1103/PhysRevC.84.054906}

\bibitem{Shen:2013vja}
C.~Shen, U.W. Heinz, J.F. Paquet, C.~Gale, Phys. Rev. C \textbf{89}(4), 044910
  (2014).
\newblock \doi{10.1103/PhysRevC.89.044910}

\bibitem{Massen:2024pnj}
O.~Massen, G.~Nijs, M.~Sas, W.~van~der Schee, R.~Snellings,   (2024)

\bibitem{DLS:1997kbk}
R.J. Porter, et~al., Phys. Rev. Lett. \textbf{79}, 1229 (1997).
\newblock \doi{10.1103/PhysRevLett.79.1229}

\bibitem{Specht:2007ez}
H.J. Specht, Nucl. Phys. A \textbf{805}, 338 (2008).
\newblock \doi{10.1016/j.nuclphysa.2008.02.275}

\bibitem{NA60:2008ctj}
R.~Arnaldi, et~al., Eur. Phys. J. C \textbf{61}, 711 (2009).
\newblock \doi{10.1140/epjc/s10052-009-0878-5}

\bibitem{Geurts:2012rv}
F.~Geurts, Nucl. Phys. A \textbf{904-905}, 217c (2013).
\newblock \doi{10.1016/j.nuclphysa.2013.01.062}

\bibitem{PHENIX:2005kfn}
K.~Ozawa, et~al., Eur. Phys. J. C \textbf{43}, 421 (2005).
\newblock \doi{10.1140/epjc/s2005-02212-3}

\bibitem{Feuillard:2022ket}
V.J.G. Feuillard, PoS \textbf{PANIC2021}, 233 (2022).
\newblock \doi{10.22323/1.380.0233}

\bibitem{CJHP2024}
C.~Jin.
\newblock {\em Hard Probes 2024} (2024)

\bibitem{Huck:2014mfa}
P.~Huck, Nucl. Phys. A \textbf{931}, 659 (2014).
\newblock \doi{10.1016/j.nuclphysa.2014.09.090}

\bibitem{Geurts:2022xmk}
F.~Geurts, R.A. Tripolt, Prog. Part. Nucl. Phys. \textbf{128}, 104004 (2023).
\newblock \doi{10.1016/j.ppnp.2022.104004}

\bibitem{Kapusta:1993hq}
J.I. Kapusta, E.V. Shuryak, Phys. Rev. D \textbf{49}, 4694 (1994).
\newblock \doi{10.1103/PhysRevD.49.4694}

\bibitem{Hohler:2013eba}
P.M. Hohler, R.~Rapp, Phys. Lett. B \textbf{731}, 103 (2014).
\newblock \doi{10.1016/j.physletb.2014.02.021}

\bibitem{NA60:2022sze}
C.~Ahdida, et~al.
\newblock {Letter of Intent: the NA60+ experiment} (2022)

\bibitem{Rapp:2014hha}
R.~Rapp, H.~van Hees, Phys. Lett. B \textbf{753}, 586 (2016).
\newblock \doi{10.1016/j.physletb.2015.12.065}

\bibitem{STAR:2024bpc}
STAR.
\newblock {Temperature Measurement of Quark-Gluon Plasma at Different Stages}
  (2024)

\bibitem{Churchill:2023zkk}
J.~Churchill, L.~Du, C.~Gale, G.~Jackson, S.~Jeon, Phys. Rev. Lett.
  \textbf{132}(17), 172301 (2024).
\newblock \doi{10.1103/PhysRevLett.132.172301}

\bibitem{Wong:1991be}
S.M.H. Wong, Z. Phys. C \textbf{53}, 465 (1992).
\newblock \doi{10.1007/BF01625907}

\bibitem{Kasmaei:2018oag}
B.S. Kasmaei, M.~Strickland, Phys. Rev. D \textbf{99}(3), 034015 (2019).
\newblock \doi{10.1103/PhysRevD.99.034015}

\bibitem{Martinez:2008di}
M.~Martinez, M.~Strickland, Phys. Rev. C \textbf{78}, 034917 (2008).
\newblock \doi{10.1103/PhysRevC.78.034917}

\bibitem{Coquet:2021lca}
M.~Coquet, X.~Du, J.Y. Ollitrault, S.~Schlichting, M.~Winn, Phys. Lett. B
  \textbf{821}, 136626 (2021).
\newblock \doi{10.1016/j.physletb.2021.136626}

\bibitem{Garcia-Montero:2024lbl}
O.~Garcia-Montero, P.~Plaschke, S.~Schlichting,   (2024)

\bibitem{Wu:2024pba}
X.Y. Wu, L.~Du, C.~Gale, S.~Jeon, Phys. Rev. C \textbf{110}(5), 054904 (2024).
\newblock \doi{10.1103/PhysRevC.110.054904}

\bibitem{JFHP2024}
J.F. Paquet.
\newblock {Conference Highlights: Electroweak Probes, {\em Hard Probes 2024}},
  and private communication (2024)

\bibitem{Steinbrecher:2018phh}
P.~Steinbrecher, Nucl. Phys. A \textbf{982}, 847 (2019).
\newblock \doi{10.1016/j.nuclphysa.2018.08.025}

\bibitem{Chatterjee:2007xk}
R.~Chatterjee, D.K. Srivastava, U.W. Heinz, C.~Gale, Phys. Rev. C \textbf{75},
  054909 (2007).
\newblock \doi{10.1103/PhysRevC.75.054909}

\bibitem{Geurts:2013kiz}
F.~Geurts, J. Phys. Conf. Ser. \textbf{458}, 012016 (2013).
\newblock \doi{10.1088/1742-6596/458/1/012016}

\bibitem{Speranza:2017hso}
E.~Speranza, {Virtual photon polarization and dilepton anisotropy in
  pion-nucleon and heavy-ion collisions}.
\newblock Ph.D. thesis, Darmstadt, Tech. U. (2018).
\newblock \doi{10.15120/GSI-2018-01176}

\bibitem{Seck:2023oyt}
F.~Seck, B.~Friman, T.~Galatyuk, H.~van Hees, R.~Rapp, E.~Speranza, J.~Wambach,
  Phys. Lett. B \textbf{861}, 139267 (2025).
\newblock \doi{10.1016/j.physletb.2025.139267}

\bibitem{NA60:2008iqj}
R.~Arnaldi, et~al., Phys. Rev. Lett. \textbf{102}, 222301 (2009).
\newblock \doi{10.1103/PhysRevLett.102.222301}

\bibitem{Baym:2017qxy}
G.~Baym, T.~Hatsuda, M.~Strickland, Phys. Rev. C \textbf{95}(4), 044907 (2017).
\newblock \doi{10.1103/PhysRevC.95.044907}

\bibitem{Coquet:2023wjk}
M.~Coquet, M.~Winn, X.~Du, J.Y. Ollitrault, S.~Schlichting, Phys. Rev. Lett.
  \textbf{132}(23), 232301 (2024).
\newblock \doi{10.1103/PhysRevLett.132.232301}

\bibitem{Hauksson:2023dwh}
S.~Hauksson, C.~Gale, Phys. Rev. C \textbf{109}(3), 034902 (2024).
\newblock \doi{10.1103/PhysRevC.109.034902}

\bibitem{Wu:2024vyc}
X.Y. Wu, H.~Gao, B.~Forster, C.~Gale, G.~Jackson, S.~Jeon, Phys. Rev. Lett.
  \textbf{134}(24), 242301 (2025).
\newblock \doi{10.1103/c4jv-jxkn}

\bibitem{NA60:2007lzy}
R.~Arnaldi, et~al., Phys. Rev. Lett. \textbf{100}, 022302 (2008).
\newblock \doi{10.1103/PhysRevLett.100.022302}

\bibitem{KumarSrivastava:1993mp}
D.~Kumar~Srivastava, C.~Gale, Phys. Lett. B \textbf{319}, 407 (1993).
\newblock \doi{10.1016/0370-2693(93)91742-6}

\bibitem{Garcia-Montero:2019kjk}
O.~Garcia-Montero, N.~L\"oher, A.~Mazeliauskas, J.~Berges, K.~Reygers, Phys.
  Rev. C \textbf{102}(2), 024915 (2020).
\newblock \doi{10.1103/PhysRevC.102.024915}

\bibitem{Shen:2016zpp}
C.~Shen, J.F. Paquet, G.S. Denicol, S.~Jeon, C.~Gale, Phys. Rev. C
  \textbf{95}(1), 014906 (2017).
\newblock \doi{10.1103/PhysRevC.95.014906}

\bibitem{Srivastava:2002ic}
D.K. Srivastava, C.~Gale, R.J. Fries, Phys. Rev. C \textbf{67}, 034903 (2003).
\newblock \doi{10.1103/PhysRevC.67.034903}

\bibitem{Yazdi:2022cuk}
R.M. Yazdi, S.~Shi, C.~Gale, S.~Jeon, Acta Phys. Polon. Supp. \textbf{16}(1), 1
  (2023).
\newblock \doi{10.5506/APhysPolBSupp.16.1-A129}

\bibitem{Bailhache:2024mck}
R.~Bailhache, et~al., Phys. Rept. \textbf{1097}, 1 (2024).
\newblock \doi{10.1016/j.physrep.2024.10.002}

\end{thebibliography}
